\documentclass[11pt,english]{article}
\usepackage[utf8]{inputenc}
\usepackage{refstyle}
\usepackage{float}
\usepackage{amsmath}
\usepackage{amssymb}
\usepackage{graphicx}
\usepackage{esint}
\makeatletter

\AtBeginDocument{\providecommand\refb[1]{(\ref{eq:#1})}}
\AtBeginDocument{\renewcommand\eqref[1]{Eq.~(\ref{eq:#1})}}
\AtBeginDocument{\providecommand\eqsref[1]{Eqs.~(\ref{eq:#1})}}
\AtBeginDocument{\renewcommand\figref[1]{Fig.~\ref{fig:#1}}}
\AtBeginDocument{\renewcommand\tabref[1]{Table~\ref{tab:#1}}}
\AtBeginDocument{\renewcommand\secref[1]{Section~\ref{sec:#1}}}
\AtBeginDocument{\providecommand\appnref[1]{Appendix~\ref{sec:#1}}}

\usepackage{jcappub}
\usepackage{slashed}
\usepackage[dvipsnames]{xcolor}
\usepackage[normalem]{ulem}
\usepackage{comment}

\DeclareMathOperator{\diag}{diag}

 \normalsize

\title{Probing Right Handed Neutrino assisted Reheating with Gravitational Waves and Leptogenesis}

\author[1]{Arghyajit Datta,}
\author[2]{Shaaban Khalil,}
\author[3]{Rajat Kumar Mandal}
\author[3]{and Arunansu Sil}

\affiliation[1]{
Department of Physics, Chung-Ang University, Seoul 06974, Korea}
\affiliation[2]{Center for Fundamental Physics,
Zewail City of Science and Technology, 6th October City 12566, Giza, Egypt}
\affiliation[3]{Department of Physics, Indian Institute of Technology Guwahati,
 Assam-781039, India}

\emailAdd{arghyad053@gmail.com}
\emailAdd{skhalil@zewailcity.edu.eg}
\emailAdd{r.mandal@iitg.ac.in}
\emailAdd{asil@iitg.ac.in}

\abstract{
We investigate a non-instantaneous reheating period in the early Universe, where the inflaton field decays exclusively to right-handed neutrinos (RHNs). 
The subsequent decay of these RHNs into Standard Model particles not only drives the transition to a radiation-dominated era but also generates the baryon asymmetry of the Universe via leptogenesis. 
In this typical reheating scenario, gravitational waves (GWs) can be produced during inflaton decay, both through bremsstrahlung and inflaton scattering processes.
While GW production via bremsstrahlung dominates near the end of the reheating phase, inflaton scattering leads to a non-negligible GW contribution near the maximum temperature of the Universe. 
The combined GW spectrum from both decay and scattering processes lies within the sensitivity range of proposed resonant cavity experiments. 
This framework thus offers a compelling and unified approach to addressing neutrino mass generation, the baryon asymmetry of the Universe via leptogenesis, and probing the dynamics of a non-instantaneous reheating era.}

\makeatother

\usepackage{babel}
\begin{document}
\begin{flushright}
CPHT-RR017.0515
\end{flushright}
\maketitle

\section{Introduction}

The successful predictions of the relative abundances of light elements during Big Bang Nucleosynthesis (BBN), assuming thermal (chemical) equilibrium at a temperature \( T_{\text{BBN}} \sim 1\,\text{MeV} \), and their remarkable agreement with observational data~\cite{Valerdi_2019, Hsyu_2020, Aver_2021, Valerdi_2021}, strongly support the existence of a radiation-dominated Universe at the time of BBN.
On the other hand, the observed anisotropies in the cosmic microwave background radiation and the large-scale structure of the Universe provide strong support for the inflationary paradigm in the very early Universe~\cite{Planck_2020_10}. 
This inflationary epoch is typically described as being driven by a scalar field, whose large vacuum energy density leads to a period of exponential expansion, ultimately ending with a cold and nearly empty Universe. 
Understanding the seemingly prolonged era between this post-inflationary cold phase and the thermal Universe at the time of BBN—commonly referred to as the extended reheating period~\cite{Chung_1999, Giudice_2001, Garcia_2020, Haque_2022, Haque_2023}—remains a significant challenge and is expected to require insights from both particle physics and cosmology~\cite{Linde_2005}.

Reheating is generally characterised as the transition period during which the energy density of a scalar field (\(\phi\)), known as the inflaton—responsible for the exponential expansion of the Universe—is transferred, after the end of inflation, into the radiation energy density of the Universe. During this period, the inflaton field is expected to decay perturbatively while oscillating around the minimum of its potential. 
The decay products then become thermalised quickly\footnote{For non-instantaneous thermalisation process, see \cite{Harigaya_2014_1, Mukaida_2016, Garcia_2018}.} (provided they possess sufficiently strong interactions among themselves), and the temperature of the plasma rises sharply, followed by a non-trivial decline during the subsequent matter-dominated phase.
It is important to note that during perturbative reheating via inflaton decay, the efficiency of energy transfer from the inflaton to the Standard Model (SM) thermal bath critically depends on the coupling between the inflaton and SM fields. In general, since the inflaton is assumed to be a gauge singlet scalar, it cannot couple directly to SM fields at tree level, except to the SM Higgs doublet through\footnote{A trilinear interaction such as \(\phi H^\dagger H\) is also possible. However, such a term can be eliminated by introducing an additional symmetry, without affecting the physical properties of the system.} an interaction term of the form \(\phi^2 H^\dagger H\).
It is known that the annihilation of inflaton through such Higgs portal interaction is insufficient to successfully reheat the Universe with the simplest form of the inflaton potential~\cite{Garcia_2020}. 
Hence, in order for the inflaton to decay into SM fermions, an effective coupling of the form \(\phi \bar{f}f\) (where \(f\) denotes a SM fermion, either a quark or a lepton) is often assumed~\cite{Garcia_2020, Barman_2023, Bernal_2024, Kanemura_2024}. 
However, such an interaction is not invariant under the SM gauge symmetries, and its realisation typically requires non-trivial higher-dimensional operators or extensions beyond the SM. 
In this work, in addition to the SM content, we include right-handed neutrinos (RHNs), which are naturally motivated by their role in generating light neutrino masses via the type-I seesaw mechanism~\cite{Minkowski_1977, Yanagida_1979_1, Yanagida_1979, GellMann_1979, Mohapatra_1980, Schechter_1980, Schechter_1982, Gellmann_2013, Datta_2021} and in explaining the baryon asymmetry of the Universe (BAU) via leptogenesis~\cite{ Fukugita_1986, Luty_1992, Pilaftsis_1997, Barbieri_1999, Nardi_2005, Nardi_2006, Abada_2006, Abada_2006_1, Blanchet_2006, Dev_2017, Datta_2021_1,  Datta_2022, Bhandari_2023}.
These RHNs, being SM singlets, can naturally have a tree-level interaction with the inflaton field of the form \( y_{\phi} \phi N N \), where \( y_{\phi} \) is a coupling constant. 
This interaction leads to the decay of the inflaton into RHNs, which subsequently decay into SM leptons and Higgs doublets via their Yukawa couplings—interactions that also participate in the type-I seesaw mechanism. 
These decays populate the SM thermal bath, thereby enabling a successful reheating of the Universe. 
We refer to this framework as \emph{RHN-assisted reheating} throughout the rest of the paper. 
This scenario is particularly appealing, as it is intrinsically linked to neutrino physics and the baryon asymmetry of the Universe (BAU)$-$both well-established indications of physics beyond the Standard Model.

It is worth noting that during the first phase of this two-stage reheating process, when the inflaton decays into a pair of right-handed neutrinos (RHNs), a single graviton can be emitted via the bremsstrahlung process due to the inevitable coupling of gravitons to RHNs \cite{Ghoshal_2023}. 
Furthermore, a similar gravitational coupling with the inflaton can lead to graviton pair production via inflaton annihilation processes~\cite{Ema_2015, Ema_2016, Ema_2020, Klose_2022, Ghiglieri_2022, Choi_2024}, which may alter the gravitational wave (GW) signal generated solely by inflaton decay. 
In this work, we investigate the resulting GW spectrum produced by both the decay and \(2 \rightarrow 2\) annihilation processes of the inflaton within the \emph{RHN-assisted reheating} scenario. 
Such a GW spectrum can, in principle, encode information about the specific reheating mechanism discussed here, especially if it falls within the sensitivity reach of ongoing or proposed GW detection experiments~\cite{Aggarwal_2021, Herman_2021, Berlin_2022, Domcke_2022, Herman_2023, Bringmann_2023, He_2023}. 
Hence, the detection of such gravitational waves would serve as a distinctive signature of an extended reheating epoch following inflation.
During the second stage of RHN-assisted reheating, as the RHNs decay and populate the radiation bath, a finite amount of \(B-L\) asymmetry is generated due to the Majorana nature of the RHNs. 
By requiring that this generated asymmetry accounts for the observed baryon asymmetry of the Universe, we impose constraints on the reheating dynamics and subsequently discuss their implications for the predicted GW spectrum.

Previous studies have investigated GW production during reheating under various assumptions regarding the inflaton's couplings to other fields and the processes contributing to GW generation.
Some works have focused exclusively on inflaton decay, without considering annihilation processes~\cite{Nakayama_2019, Barman_2023, Kanemura_2024}, while others have examined only inflaton annihilation or employed simplified reheating scenarios in which the reheating temperature (\(T_{\text{RH}}\)) is treated as a free parameter~\cite{Choi_2024, Bernal_2024}, or assumed instantaneous reheating~\cite{Ghoshal_2023}. 
Although Ref.~\cite{Bernal_2024} studies GW production from both inflaton decay and \(2 \rightarrow 3\) annihilation, the reheating mechanism still relies on an effective coupling between the inflaton and SM fields.
In contrast, we propose a renormalisable interaction between the inflaton field and beyond-the-Standard-Model (BSM) singlet right-handed neutrinos (RHNs), whose subsequent decays are responsible for reheating as well as the generation of the baryon asymmetry of the Universe (BAU).
We consider both the decay and annihilation of the inflaton field as sources of GW production. 
In this framework, the reheating temperature is not a free parameter but emerges dynamically from the underlying reheating process, with parameters constrained by the requirement of successful leptogenesis. 
This, in turn, impacts the spectral characteristics of the resulting primordial GWs.
Thus, the present setup provides a more comprehensive and self-consistent framework for gravitational wave production during perturbative extended reheating, while simultaneously accounting for the origin of neutrino masses and the observed baryon asymmetry of the Universe via leptogenesis.

The paper is structured as follows: \secref{sec-2} explores the era of extended reheating with right-handed neutrinos (RHNs), beginning with the Boltzmann equations for the inflaton, RHNs, and radiation, followed by an estimation of the reheating temperature within the context of RHN-assisted reheating. \secref{sec-3} presents the formalism for gravitational wave (GW) generation during RHN-assisted reheating, covering GW production from both inflaton decay and annihilation processes, along with the resulting GW spectrum and its connection to leptogenesis. In \secref{sec4}, we investigate the combined phenomenology of the GW spectrum and leptogenesis. Finally, our conclusions and main findings are summarised in \secref{sec5}.

\section{The Era of Extended Reheating with RHNs}
\label{sec:sec-2}
As the extended era of reheating plays a central role in this study, we begin by outlining the framework of RHN-assisted reheating. RHNs are among the most natural extensions of the SM, particularly in explaining the smallness of light neutrino masses via the type-I seesaw mechanism. In this work, we further extend their interactions to include a coupling with the inflaton field $\phi$ through the term $y_{\phi} \phi NN$. This minimal setup enables the Universe to reheat in two distinct phases.
First, after the end of the inflationary era, the inflaton field decays into RHNs. The subsequent out-of-equilibrium decay of these RHNs into SM lepton and Higgs doublets not only generates the thermal bath but also accounts for the BAU via leptogenesis. To elucidate the discussion on reheating in this section, we mention here only the relevant action for reheating (discussion on generation of GW and corresponding Lagrangian following the subsequent sections) involving $\phi, N$ and the SM fields as 
\begin{equation}
 S_M = \intop d^{4}x\sqrt{-g}\left(
    \mathcal{L}_{\phi}^{\text{kin}}-V(\phi)+\mathcal{L}_{N}^{\text{kin}}+\mathcal{L}_{\text{SM}}+\mathcal{L}_{\text{int}}\right).\label{eq:actionfunctional-M}
\end{equation}
Here, $\mathcal{L}_{\phi}^{\text{kin}}$ and $\mathcal{L}_{N}^{\text{kin}}$ are kinetic parts of the inflaton $\phi$ and RHN field $N$ respectively while $\mathcal{L}_{\text{int}}$ contains the type-I seesaw as well as the inflaton-RHN interaction as expressed by, 
\begin{equation}
\mathcal{L}_{\text{int}}\supset\left( Y_{\nu}\right)_{\alpha i}\bar{\ell}_{L_\alpha}\tilde{H}N_i+\frac{1}{2}M_{i}\overline{N^{c}_i}N_i+ y_{\phi}\phi\overline{N^{c}_i}N_i,\label{eq:interactionlagrangian}
\end{equation}
where \( \ell_{L}^{T} = (\nu_{L}, e_{L}) \) and \( H \) denote the left-handed SM lepton and Higgs doublets, respectively.
Moreover, \( y_\phi \) represents the coupling constant that governs the strength of the interaction between the inflaton and RHNs, and accordingly determines the reheating temperature \( T_{\text{RH}} \). While there is no direct evidence pinpointing the exact onset of the radiation-dominated era, it is well established that by the time of BBN, the Universe must have been in a radiation-dominated state. This uncertainty in the precise value of \( T_{\text{RH}} \) motivates treating \( y_\phi \) as a free parameter in our analysis, subject to the lower bound \( T_{\text{RH}} \gtrsim ~\text{few MeV} \)~\cite{Kawasaki:1999na,Hasegawa:2019jsa}. We elaborate on the inflaton potential \( V(\phi) \) in the context of reheating below.

Note that the end of inflation is marked by the onset of oscillations of the inflaton field around the origin. These oscillations are typically assumed to be harmonic; however, any deviation from harmonicity (i.e., anharmonic effects) can non-trivially impact the reheating dynamics~\cite{ Bernal_2019, Garcia_2020}. 
In what follows, we adopt the formalism developed in~\cite{Garcia_2021}, which allows for both harmonic and anharmonic scenarios. These are closely related to the limiting behaviour of the inflaton potential near its minimum, where the field amplitude is small. In this regime, we consider a generic power-law potential of the form
\begin{equation}
	V(\phi) = \lambda \frac{|\phi|^k}{M_P^{k-4}}. \label{eq:potentialinflaton}
\end{equation}
Such a potential can arise naturally in T-attractor models\footnote{For $k = 2$, the potential corresponds to that of Starobinsky-type inflation \cite{Starobinsky_1980} in the large-field limit.}~\cite{Ellis_2020, Sloan_2020}, particularly within the framework of no-scale supergravity~\cite{Kallosh_2013, Khalil_2019}.

\subsection{Evolution of Energy Densities during Reheating} 
\label{sec:envelope-approximation}

To describe the evolution of the inflaton field during the reheating era, we primarily adopt a semi-classical approach. In this framework, the inflaton is treated as a classical condensate, with its action extracted from \eqref{actionfunctional-M} given by
\begin{align}
S[{\phi}]=\intop d^{4}x\sqrt{-g}\left(\frac{1}{2}\partial_{\mu}\phi\partial^{\mu}\phi-V\left(\phi\right)\right),\label{eq:actionphi}
\end{align}
minimisation of which leads to the equation of motion for the inflaton field
\begin{equation}
\ddot{\phi}+3{\mathcal{H}}\dot{\phi}+V'(\phi)=0, \label{eq:eomphi}
\end{equation}
where $\mathcal{H}$ is the Hubble parameter. 
The time dependence of the inflaton as a solution of this equation can be parametrised by \cite{Garcia_2021} 
\begin{equation}
\phi(t)=\phi_{0}(t)\cdot\mathcal{P}(t),\label{eq:envelopeapproximation}
\end{equation}
where the amplitude $\phi_{0}$ (the envelope function) captures the effects of redshift (and decay also, which we elaborate upon soon) while $\mathcal{P}$ denotes the effects of (an)harmonicity of the short timescale oscillations.
Since the factorisation in \eqref{envelopeapproximation} is not unique, i.e., $\phi_{0}\mathcal{P}=\left(\phi_{0}/c\right)\left(c\mathcal{P}\right)$ upto a constant $c$, one needs to fix a definition of either $\phi_{0}$ or $\mathcal{P}$.
A convenient definition of $\phi_{0}$ can be taken as \cite{Garcia_2020} 
\begin{equation}
V\left(\phi_{0}\right)=\left\langle \rho_{\phi}\left(t\right)\right\rangle, \label{eq:defenvelope}
\end{equation}
where $\left\langle \ldots\right\rangle $ denotes average over one oscillation. 
The inflaton energy density $\rho_{\phi}\left(t\right)$ and additionally, the inflaton pressure $P_{\phi}\left(t\right)$ are obtained (using the stress energy tensor $T^{\mu\nu}\left(\phi\right)$) as, 
\begin{align}
\rho_{\phi}\left(t\right) & =\frac{1}{2}\dot{\phi}^{2}+V\left(\phi\right)\label{eq:energydensityinflaton},\\
P_{\phi}\left(t\right) & =\frac{1}{2}\dot{\phi}^{2}-V\left(\phi\right).\label{eq:pressureinflaton}
\end{align}
Note that taking average over one oscillation period of \eqsref{energydensityinflaton} and \refb{pressureinflaton} results in a barotropic relation of the form:
\begin{equation}
\left\langle P_{\phi}\right\rangle =w_{\phi}\left\langle \rho_{\phi}\right\rangle, \quad\text{where}\quad w_{\phi}=\frac{k-2}{k+2}.\label{eq:eosinflaton}
\end{equation}
Furthermore, using \eqsref{potentialinflaton}, \refb{defenvelope} and the definition of inflaton mass $m_{\phi}^{2}\left(t\right)=V''\left(\phi_{0}\right)$, we get 
\begin{align}
m_{\phi}\left(t\right) & =\sqrt{\frac{k\left(k-1\right)}{\phi_{0}^{2}}\langle\rho_{\phi}\rangle},\label{eq:massinflaton1}\\
& =\sqrt{k\left(k-1\right)}M_{P}^{\frac{4-k}{k}}\lambda^{\frac{1}{k}}{\langle\rho_{\phi}\rangle}^{\frac{k-2}{2k}}.\label{eq:massinflaton}
\end{align}
Now to determine the periodicity of the oscillatory part $\mathcal{P}\left(t\right)$, we substitute \eqsref{potentialinflaton}, \refb{envelopeapproximation}, and \refb{massinflaton1} into \refb{energydensityinflaton}.
Since the envelope $\phi_{0}$ varies slowly (related to the Hubble parameter and the decay width of the inflaton), \emph{i.e.}, $\dot{\phi_{0}} \simeq 0$ over one oscillation, \eqref{energydensityinflaton} leads to
\begin{equation}
\dot{\mathcal{P}}(t)^{2}=\frac{2m_{\phi}^{2}\left(t\right)}{k\left(k-1\right)}\left(1-\mathcal{P}(t)^{k}\right), \label{eq:P-eq}
\end{equation}
which, upon integrating between the turning points $\mathcal{P}=\pm1$, determines the angular frequency of the oscillation as
\begin{equation}
\omega\left(t\right)=m_{\phi}\left(t\right)\sqrt{\frac{\pi k}{2\left(k-1\right)}}\frac{\Gamma\left(\frac{1}{2}+\frac{1}{k}\right)}{\Gamma\left(\frac{1}{k}\right)}. \label{eq:freqinflaton}
\end{equation}
The oscillatory part $\mathcal{P}\left(t\right)$ can then be expanded into its Fourier series with respect to this frequency as
\begin{equation}
\mathcal{P}\left(t\right)=\sum_{n=-\infty}^{\infty}\mathcal{P}_{n}e^{-in\omega t}. \label{eq:P-expansion}
\end{equation}

To study the effect of redshift over longer timescales, we average \eqref{eomphi} over one oscillation period. Substituting \eqref{defenvelope} and \eqref{eosinflaton} into \eqref{eomphi}, we obtain the evolution equation for the energy density of the $\phi$ field:
\footnote{Henceforth, we omit the $\left\langle \cdots \right\rangle$ notation for oscillation averages, and adopt this convention implicitly.}
 \begin{equation}
 	\dot{\rho}_{\phi} + 3\mathcal{H}(1 + w_{\phi})\rho_{\phi} = 0. \label{eq:BTEinflaton-1}
 \end{equation}
Additionally, to account for the production of right-handed neutrinos (RHNs) from inflaton decay, induced by the coupling $y_{\phi}\phi\,\overline{N^{c}}N$ in the oscillating background, one must modify the right-hand side of \eqref{BTEinflaton-1} to include a collision term. This yields:
\begin{equation}
	\dot{\rho}_{\phi} + 3\mathcal{H}(1 + w_{\phi})\rho_{\phi} = C[f_{\phi}] \equiv - (1 + w_{\phi}) \Gamma_{\phi} \rho_{\phi}, \label{eq:BTEinflaton}
\end{equation}
where $C[f_{\phi}]$ represents the energy transfer due to the decay of the inflaton into RHNs, characterised by the decay width $\Gamma_{\phi}$.

Given the interaction Lagrangian density  in \eqref{interactionlagrangian}, the Feynman amplitude of this process for the $n$th oscillation mode of inflaton can be evaluated as
\begin{equation}
\mathcal{M}_{n}=y_{\phi}\phi_{0}\mathcal{P}_{n}\bar{u}\left(p_{2}\right)v\left(p_{3}\right).
\end{equation}
 The collision integral then leads to
$C\left[f_{\phi}\right]=\sum_{n}\intop E_{n}\left|\mathcal{M}_{n}\right|^{2}D_{2},
$
where $E_{n}=n\omega\left(t\right)$ is the energy of $n$th mode of inflaton oscillation and $D_{2}$ is the two-body phase space of the outgoing RHNs. 
After taking average over one oscillation period, $C\left[f_{\phi}\right]$ takes the form
\begin{equation}
C\left[f_{\phi}\right]=\frac{2k\left(k-1\right)}{8\pi}y_{\phi}^{2}\,\rho_{\phi}(t)\,\omega\left(t\right)\left[\frac{\omega\left(t\right)}{m_{\phi}\left(t\right)}\right]^{3}\left[\sum_{n}n^{3}\left|\mathcal{P}_{n}\right|^{2}\left(1-\frac{4M_{N_{i}}^{2}(t)}{n^{2}\omega^{2}(t)}\right)^{\frac{3}{2}}\right],\label{eq:decaywidthinflaton2}
\end{equation}
where $M_{N_i}(t)$ includes the bare contribution and the field dependent (due to its interaction with inflaton) contribution to the RHN mass, structure of which can be evaluated as 
\begin{equation}
M_{N_{i}}\left(t\right)=M_{i}+M_{i}(\phi),~\text{with}~~ M_{i}(\phi)= y_{\phi}M_{P}^{\frac{k-4}{k}}\left[\frac{\rho_{\phi}\left(t\right)}{\lambda}\right]^{\frac{1}{k}}.\label{eq:massRHN}
\end{equation}
Note that this field dependent correction $M_i(\phi)$ becomes important earlier in the reheating process when $\rho_{\phi}$ is large.
 \eqref{BTEinflaton} can now be solved  given the parameters involved \emph{i.e.}, $y_{\phi}, M_{i}, \lambda$ with a specific choice of $k$ while information of $\mathcal{P}$ is to be obtained using \eqref{P-eq} in conjunction with \eqref{P-expansion}.

As the inflaton decays into RHNs, the evolution of the number density of the RHNs can now be expressed by the Boltzmann equation
\begin{equation}
\dot{n}_{N_{i}}+3\mathcal{H}n_{N_{i}}=\Gamma_{\phi}\frac{\rho_{\phi}}{m_{\phi}}-\Gamma_{N_{i}}(n_{N_{i}}-n_{N_{i}}^{\text{eq}}),\label{eq:BTERHN}
\end{equation}
where $\Gamma_{N_{i}}=\frac{(Y_{\nu}^{\dagger}Y_{\nu})_{ii}}{8\pi}M_{N_{i}}(t)$ corresponds to the decay rate of the RHN $N_i$ decaying to SM lepton and Higgs doublets (imitating as the radiation) and $n_{N_{i}}^{\text{eq}}$ is the equilibrium number density of the RHN. 
The analogous equation for the radiation energy density would then take the form
\begin{equation}
\dot{\rho}_{R}+4\mathcal{H}\rho_{R}=\Gamma_{N_{i}}(n_{N_{i}}-n_{N_{i}}^{{\rm eq}})E_{N_{i}}, \label{eq:BTEradiaion}
\end{equation}
where $E_{N_{i}}$ is the energy of a single RHN produced from inflaton decay, given by
\begin{equation}
E_{N_{i}} = \sqrt{ M_{N_{i}}^{2}(t) + \frac{1}{\mathcal{A}^{2}} \left[ \frac{m_{\phi}^{2}}{4} - M_{N_{i}}^{2}(t) \right] }, \label{eq:energyRHN}
\end{equation}
with $\mathcal{A} = a/a_{\text{end}}$ being the scale factor normalised to its value at the end of inflation.

\subsection{Estimating Reheating Temperature Embedded in RHN-Assisted Reheating}
\label{sec:reheating-mechanism}

To learn about the exact reheating process, one now needs to solve the following equations (as obtained above in \eqsref{BTEinflaton}, \refb{BTERHN} and \refb{BTEradiaion} simultaneously 
\begin{align}
	\dot{\rho}_{\phi} + 3\mathcal{H}(1 + w_{\phi}) \rho_{\phi} &= -\Gamma_{\phi}(1 + w_{\phi}) \rho_{\phi}, \label{eq:phi} \\
	\dot{n}_{N_{i}} + 3\mathcal{H} n_{N_{i}} &= \Gamma_{\phi} \frac{\rho_{\phi}}{m_{\phi}} - \Gamma_{N_{i}} (n_{N_{i}} - n_{N_{i}}^{\text{eq}}), \label{eq:N} \\
	\dot{\rho}_{R} + 4\mathcal{H} \rho_{R} &= \Gamma_{N_{i}} (n_{N_{i}} - n_{N_{i}}^{\text{eq}}) E_{N_{i}}, \label{eq:r}
\end{align}
where the Hubble expansion rate is given by, 
\begin{equation}
\mathcal{H}^{2}=\frac{\rho_{\phi}+\sum_i n_{N_{i}}E_{N_{i}}+\rho_{R}}{3M_{p}^{2}}.\label{eq:parameterhubble}
\end{equation}

\eqref{phi} describes the evolution of the inflaton energy density, including its decay into RHNs and other species.  
\eqref{N} governs the number density of RHNs, accounting for their production from inflaton decay and their decay toward thermal equilibrium.  
Finally, \eqref{r} captures the evolution of the radiation energy density, sourced by RHN decays.

Assuming instantaneous thermalisation of the produced SM particles from the $N_i$ decays, the temperature of the radiation bath thus produced is governed by 
\begin{equation}
T=\alpha^{-\frac{1}{4}}\rho_{R}^{\frac{1}{4}}; ~~{\rm with}\quad\alpha=\frac{\pi^{2}}{30}g\left(T\right), \label{eq:def-temperature}
\end{equation}
where $g\left(T\right)$ is the total degrees of freedom of all the constituent fields of the radiation bath.
Together, these equations allow us to evaluate the reheating temperature and also aid in estimating the temporal behaviour of different components of the Universe as it transits from an inflaton-dominated phase to a radiation-dominated one.

Solving \eqref{phi}– \eqref{r} simultaneously requires knowledge of the inflaton energy density $\rho_{\phi}^{\text{end}}$ at the end of inflation, which is characterised by the vanishing of cosmic acceleration, i.e., $\ddot{a} = 0$. This condition, combined with the Friedmann equation, leads to $\rho_{\phi}^{\text{end}}=\frac{3}{2}V\left(\phi_{\text{end}}\right)$. 
The exact value of $\phi_{\text{end}}$ depends on the shape of the inflaton potential away from its minimum and the detailed mechanism through which inflation ends. However, a useful approximation is given by the condition that the first slow-roll parameter reaches unity, i.e., $\epsilon = 1$, where $\epsilon=\frac{M_{P}^{2}}{2}\left(\frac{V'\left(\phi\right)}{V\left(\phi\right)}\right)^{2}$. 
For the T-attractor potentials of the form 
\begin{equation}
V\left(\phi\right)=\lambda M_{p}^{4} \left[\sqrt{6} \tanh{\left(\frac{\phi}{\sqrt{6}M_{p}}\right)}\right]^k,
\end{equation}
this yields the following expression for the inflaton field value at the end of inflation:
\begin{equation}
\phi_{\text{end}}\simeq\sqrt{\frac{3}{8}}M_{P}\ln{\left(\frac{1}{2}+\frac{k}{3}\left(k+\sqrt{k^{2}+3}\right)\right)}.
\end{equation}
Note that the reheating dynamics in this case depends on three quantities with a specific choice of $k$: $y_{\phi}$, $M_{N_i}(t)$ and $Y_{\nu}$ (apart from the parameter\footnote{\label{Footnote:lambda-calc} The inflationary parameter $\lambda$ is determined from the CMB observation of scalar perturbation spectrum amplitude $A_{s}=\frac{V\left(\phi_{*}\right)}{24\pi^{2}M_{P}^{4}\epsilon\left(\phi_{*}\right)}=2.1\times 10^{-9}$
at the Planck pivot scale $q_{*}=0.05\text{ Mpc}^{-1}$, for a fixed number of e-folds $N_*$ between $\phi_*$ and $\phi_{\rm end}$.} $\lambda$).
Among them, $y_{\phi}$ and $M_{N_i}(t)$ fix the decay rate of the inflaton via \eqref{decaywidthinflaton2}. 
On the other hand, $Y_{\nu}$ and $M_{N_i}(t)$ determine the decay rate of the RHN. 
From a minimality point of view, we consider only two RHNs ($N_{1,2}$) which are coupled to the inflaton via the same $y_{\phi}$ coupling in this setup. 
Moreover, the mass hierarchy between these two RHNs ($M_{1}\ll M_{2}$) is so chosen that $N_{2}$ remains heavier than the inflaton. 
This indicates that the lightest RHN $N_{1}$ (after being produced from the inflaton decay) will be the sole source of radiation production and lepton asymmetry generation. 
Hence, we need to solve the Boltzmann equation \eqref{BTERHN} for $N_1$ and contributions to the decay rates from other RHN drop out. 
The Yukawa coupling $Y_{\nu}$ in such a minimal scenario can be evaluated using the well-known Casas-Ibarra (CI) parametrisation, which we outline in \secref{numeric-leptogenesis}. 
Finally, the reheating temperature $T_{\rm RH}$ in this setup can be extracted from the condition: (1) $\rho_\phi=\rho_R$, if inflaton energy density dominates during the decay of RHN, or (2) $\rho_{N_i}= M_{i} n_{N_i}= \rho_R$ when there is a domination of RHN energy density after the end of inflaton domination.
We now proceed to the discussion of GW generation during such a reheating scenario.

\section{Gravitational Wave Generation during RHN-Assisted Reheating}
\label{sec:sec-3}

\begin{figure}[h]
\includegraphics[width=0.25\textwidth]{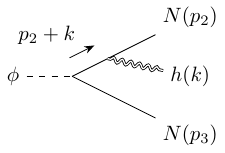}\includegraphics[width=0.25\textwidth]{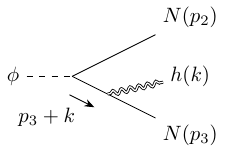}\includegraphics[width=0.25\textwidth]{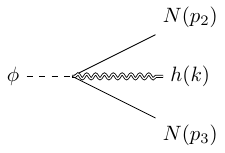}\includegraphics[width=0.25\textwidth]{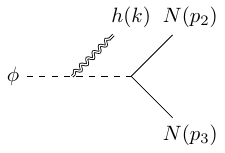} 
\caption{Feynman diagram for graviton production from three body decay of inflaton.} \label{fig:Feynman-diagram-for-decay}
\end{figure}

GWs can be associated to the decay of the inflaton field, which played a pivotal role in the reheating phase following the primordial inflation. 
As the inflaton decays to RHNs, GWs may also be generated through mechanism such as bremsstrahlung~\cite{Nakayama_2019, Barman_2023, Hu_2024_1, Hu_2024, Kanemura_2024, Datta_2024, Bernal_2024_1, Choi_2025, Konar_2025, Murayama_2025} as illustrated in the Feynman diagrams of \figref{Feynman-diagram-for-decay}. 
These waves carry signatures of the inflaton's dynamics after inflation and its interactions with other particles, making their detection a powerful tool for probing the reheating aspect of the early Universe. 
It turns out that the GWs can also be produced from inflaton annihilation during the period of reheating. 
In fact, near the onset of inflaton oscillation where inflaton number density is significantly high, the inflaton annihilation producing GWs can be the dominant source of gravitational waves, while at a later stage of the reheating, decay of the inflaton is expected to dominate the gravitational wave spectrum.

The Einstein-Hilbert action functional responsible for both the GW production mechanisms is given by 
\begin{align}
    S= S_M + \intop d^{4}x\sqrt{-g} \left( 2\kappa^{-2}\mathcal{R}\right),\label{eq:actionfunctional-full}
\end{align}
where $\kappa=\frac{2}{M_{P}}$ with $M_{P}=2.8\times10^{18}\text{ GeV}$ denoting the reduced Planck mass. 
Here, $\mathcal{R}$ denotes the Ricci scalar and $S_M$ is the matter action functional presented in Eq. \ref{eq:actionfunctional-M}.
Within the weak field limit, the metric $g_{\mu\nu}$ therein can be perturbed as $g_{\mu\nu}=\eta_{\mu\nu}+\kappa h_{\mu\nu}$ around the flat Minkowski metric $\eta_{\mu\nu}$. This additional term in \eqref{actionfunctional-full} then describes graviton propagation and self-interactions. 
Subsequently, couplings with canonically normalised graviton $h_{\mu\nu}$ and stress-energy tensors $T_{\text{SM}}^{\mu\nu}$, $T_{X}^{\mu\nu}$ (with $X$ denoting the inflaton and RHN) emerge from the kinetic terms in the matter action $S_{M}$ at the first order in $\kappa$, which takes the form \cite{Choi_1995,Holstein_2006}
\begin{align}
\mathcal{L}_{h}=\frac{\kappa}{2}h_{\mu\nu}\left(T_{\text{SM}}^{\mu\nu}+\sum_{X=\phi,N_i} T_{X}^{\mu\nu}\right).\label{eq:tmunu}
\end{align}
Here, the stress-energy tensors for a fermion ($\psi$) and a scalar ($\phi$) in the flat background metric are given by 
\begin{align}
T_{\psi}^{\mu\nu} & =\frac{i}{4}\left(\bar{\psi}\gamma^{\mu}\partial^{\nu}\psi+\bar{\psi}\gamma^{\nu}\partial^{\mu}\psi\right)-\eta^{\mu\nu}\left(\frac{i}{2}\bar{\psi}\gamma^{\alpha}\partial_{\alpha}\psi-m_{\psi}\bar{\psi}\psi\right),\label{eq:coupling-inflaton-1}\\
T_{\phi}^{\mu\nu} & =\partial^{\mu}\phi\partial^{\nu}\phi-\eta^{\mu\nu}\left(\frac{1}{2}\partial^{\alpha}\phi\partial_{\alpha}\phi-V\left(\phi\right)\right).\label{eq:coupling-RHN-1}
\end{align}
Expanding $S_{M}$ to the order of $\kappa^2$ similarly leads to further interaction terms associated with gravitons that are responsible for GW production from inflaton decay and annihilation.
All these interactions of order $\mathcal{O}\left(\kappa\right)$ and $\mathcal{O}\left(\kappa^{2}\right)$ with their corresponding Feynman diagrams and vertices are provided in \appnref{Feynman-rules}.

{ In this analysis, we consider the gravitons described by the field $h_{\mu \nu}$ as massless, spin-2 particles. 
To effectively describe the dynamics of these particles, we must impose additional gauge conditions on $h_{\mu\nu}$. 
These conditions are essential since they will reduce the total degrees of freedom to the two polarisation states of the graviton. 
We have chosen the transverse, traceless gauge in which the graviton polarisation tensors $\epsilon^{\mu\nu}\left(k,\sigma\right)$ satisfy
\begin{align}
k_{\mu}\epsilon^{\mu\nu}\left(k,\sigma\right) & =0\label{eq:conditiontransverse}\\
\eta_{\mu\nu}\epsilon^{\mu\nu}\left(k,\sigma\right) & =0,\label{eq:conditiontraceless}
\end{align}
whereas the symmetry of the metric tensor $g_{\mu\nu}$ and the orthonormality of the two polarisation modes of the graviton give rise to the conditions
\begin{align}
\epsilon^{\mu\nu} & =\epsilon^{\nu\mu}\label{eq:conditionsymmetric}\\
\epsilon^{\mu\nu}\left(k,\sigma_{1}\right)\epsilon_{\mu\nu}\left(k,\sigma_{2}\right)^{*} & =\delta^{\sigma_{1}\sigma_{2}}\label{eq:conditionorthogonal}
\end{align}
Here $k$ denotes momentum and $\sigma$ denotes polarisation of the emitted gravitons. 
We also use the polarisation sum in this gauge as \cite{Choi_1995}
\begin{equation}
\sum_{\sigma}\epsilon^{\mu\nu}\left(k,\sigma\right)\epsilon^{\alpha\beta}\left(k,\sigma\right)^{*}=\frac{1}{2}\left(\hat{\eta}^{\mu\alpha}\hat{\eta}^{\nu\beta}+\hat{\eta}^{\mu\beta}\hat{\eta}^{\nu\alpha}-\hat{\eta}^{\mu\nu}\hat{\eta}^{\alpha\beta}\right),\label{eq:polsumgraviton}
\end{equation}
where 
\[
\hat{\eta}^{\mu\nu}=\eta^{\mu\nu}-\frac{k^{\mu}\bar{k}^{\nu}+k^{\nu}\bar{k}^{\mu}}{k\cdot\bar{k}},\quad k=\left(k^{0},\vec{k}\right),\quad\bar{k}=\left(k^{0},-\vec{k}\right).
\]

\subsection{Production of GWs from Decay of the Inflaton}
\label{sec:gw-inflaton-decay}

Coupling of the RHN-inflaton system with the graviton in \eqref{tmunu}  leads to graviton bremsstrahlung\footnote{Note that the source for the third diagram of  \figref{Feynman-diagram-for-decay}  is the inflaton-RHN Yukawa coupling $\phi NN$. }  during the production process of RHNs from inflaton decay (as illustrated in the Feynman diagrams of \figref{Feynman-diagram-for-decay}). 
In our chosen gauge, only the first two diagrams contribute\footnote{The relevant Feynman amplitudes of the diagrams are included in the appendix \ref{sec:Ampl-GWD-spectrum}.} 
to this process and the relevant differential decay rate of these decay processes can be evaluated as in \eqref{decaywidthinflatongraviton}
\begin{equation}
\frac{d\Gamma^{1\rightarrow3}}{dE_{k}}=\frac{\left(k+2\right)\left(k-1\right)}{64\pi^{3}}y_{\phi}^{2}\left(\frac{m_{\phi}\left(t\right)}{M_{P}}\right)^{2}\left(\frac{\omega\left(t\right)}{m_{\phi}\left(t\right)}\right)^{4}\sum_{n=1}^{\infty}n^{4}\left|\mathcal{P}_{n}\right|^{2}\frac{1-2x_{n}}{x_{n}}\left(2x_{n}\left(x_{n}-1\right)+1\right), \label{eq:diffdecaywidth}
\end{equation}
where $x_{n}=\frac{E_{k}}{n m_{\phi}}$ with $E_k$ denoting the energy of the single graviton and $\omega\left(t\right)$ is defined in (\ref{eq:freqinflaton}).
 The kinematics of the 3-body decay constrains the graviton energy between $0<x_{n}\leq\frac{1}{2}-2\left(\frac{M_{N_i}(t)}{m_{\phi}}\right)^2$.

Now, to learn about the production and subsequent evolution of the produced gravitons from such decay process one needs to solve the differential Boltzmann equation for the energy density of the graviton, which is given by \cite{Bernal_2024}
\begin{equation}
\frac{d}{dt}\left(\frac{d\rho_{\text{GW}}^{\rm D}}{dE_{k}}\right)+4\mathcal{H}\left(\frac{d\rho_{\text{GW}}^{\rm D}}{dE_{k}}\right)=\frac{E_{k}\rho_{\phi}}{m_{\phi}}\frac{d\Gamma^{1\rightarrow3}}{dE_{k}} .\label{eq:BTEGWD}
\end{equation}
Subsequently, to obtain the present energy density of the gravitational wave today, we first solve \eqsref{BTEinflaton}, \refb{BTERHN}, \refb{BTEradiaion} along with \refb{BTEGWD} simultaneously till the end of reheating (marked by the temperature $T_{\rm RH}$) and thereafter, evaluate the corresponding redshifts for $\rho_{\rm GW}^{\rm D}$ and $E_{k}$ which scales as $(a_{0}/a)^{4}$ and $a_{0}/a$ respectively, where $a_{0}$ is the scale factor of the Universe today.
Finally, the normalised energy density of the gravitational wave can be evaluated as 
\begin{equation}
\Omega_{\rm GW}^{\rm D} h^{2}=\frac{d\rho_{\rm GW}^{\rm D}}{d\ln{f}}\bigg|_{\rm RH}\left(\frac{h^{2}}{\rho_{c, 0}}\right)\left(\frac{\xi T_{0}}{T_{\text{RH}}}\right)^{4},\label{eq:omegaGWdecay}
\end{equation}
where we have identified energy of GW $E_{k}\left(a\right)=f\left(a\right)$ as also being the frequency of the GW.
This frequency is further related to the current day's frequency as\footnote{Hereafter frequency $f$ is not explicitly shown as a function of scale factor denotes frequency observed (redshifted to) today \emph{i.e.} $f\left(a_0\right)$ .}:
\begin{align}
	f\left(a_0\right)= \left(\frac{a}{a_{\rm RH}}\right)\left(\frac{a_{\rm RH}}{a_0}\right) f\left(a\right).
	\label{eq:f0}
\end{align}
Here $\rho_{c, 0}=8.1\times{10}^{-41} h^{2}$, $T_{0}=6.626\times{10}^{-13}\text{ GeV}$ are the critical energy density and present temperature of the Universe respectively. 
We have also defined $\xi=\left(\frac{g_{0}}{g_{\rm RH}}\right)^{\frac{1}{3}}$ and $g_{\rm RH}$, $g_{0}$ as the relativistic degrees of freedom at reheating and today respectively.

\subsection{Production of GWs from Inflaton Annihilation}
\label{sec:gw-inflaton-annihilation}

The other possibility to generate gravitational waves during reheating is through the annihilation of inflatons.
In this mechanism, after the inflation ends, the oscillating inflaton condensate self-annihilates to produce gravitons through the processes shown in \figref{Feynman-diagrams-for-annihilation}. 
 The emission of such GWs is expected to be dominant at the beginning of the reheating period when the inflaton energy density is at its maximum and to stop when the inflaton energy density decreases sharply, marking the end of reheating. 
From this point onward, the resulting spectrum of gravitational waves redshifts due to the expansion of the Universe.

The relevant Feynman rules for inflaton-graviton coupling which generate the inflaton annihilation to GWs are given in \appnref{Feynman-rules}. 
The first inflaton-graviton vertex \refb{feynrulesphigrav1} is responsible for the $s$, $t$, and $u$-channel processes of inflaton-inflaton annihilation in \figref{Feynman-diagrams-for-annihilation}, while the second inflaton-graviton vertex \refb{feynrulesphigrav2} adds another contact interaction process for the inflaton annihilation (see the last diagram of \figref{Feynman-diagrams-for-annihilation}).
\begin{figure}[h]
\includegraphics[width=0.5\textwidth]{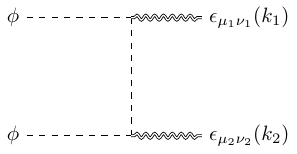}\includegraphics[width=0.5\textwidth]{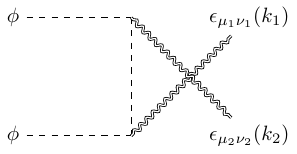}
\includegraphics[width=0.5\textwidth]{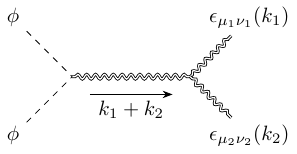}\includegraphics[width=0.5\textwidth]{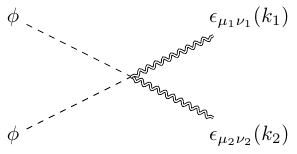}\caption{Feynman diagrams for inflaton annihilation}\label{fig:Feynman-diagrams-for-annihilation}
\end{figure}
To calculate the rate of the annihilation reaction, we need the squared Feynman amplitude for these processes. 
In \appnref{Ampl-GWS-spectrum}, we detail this calculation in the centre of momentum frame of the two outgoing gravitons while treating the incoming inflaton field as a classical condensate. 
This Feynman amplitude is oscillatory because of the inflaton oscillation and hence, we should instead consider the squared amplitude averaged over the time period of one oscillation. 
This averaging recasts the total squared amplitude as sum of squared amplitudes corresponding to the oscillation modes of the inflaton at frequencies $E_{n}=n\omega$.
Note that the Feynman amplitude for the first two diagrams in \figref{Feynman-diagrams-for-annihilation} are zero whereas the rest of the diagrams give rise to the unpolarised squared Feynman amplitude for $n$th oscillation mode 
\begin{equation}
    \left|M_{n}\right|^{2}=\frac{\phi_{0}^{4}}{2M_{p}^{4}}\widetilde{m}_{\phi}^{4}\left|\mathcal{P}^{k}_{n}\right|^{2},\quad\widetilde{m}_{\phi}^{2}=\frac{2m_{\phi}^{2}}{k\left(k-1\right)}.
\end{equation}

\subsection{Estimating Effective Gravitational Wave Spectrum}
\label{sec:gw-combined}

After the production of such GWs from inflaton annihilation, as in the case of inflaton decay, their evolution is described by a corresponding Boltzmann equation. 
The Boltzmann equation for the evolution of GW energy density $\rho_{\text{GW}}^{\rm ann.}$ in the case of annihilation is
\begin{align}
    \dot{\rho}_{\text{GW}}^{\rm ann.}+4\mathcal{H}\rho_{\text{GW}}^{\rm ann.}=\int\frac{d^{3}\vec{k}_{\text{GW}}}{(2\pi)^{3}}C\left[f_{\text{GW}}\right],\label{eq:BTE-GWS-1}
\end{align}
where the RHS evaluates to
\begin{align}
    \int\frac{d^{3}\vec{k}_{\text{GW}}}{(2\pi)^{3}}C\left[f_{\text{GW}}\right]=\left(\Delta N\right)\sum_{n}\int E_{\text{GW}}\left|M_{n}\right|^{2}D_{2},\label{eq:BTE-GWS-2}
\end{align}
as the collision term. 
Here $k_{\text{GW}}$ is the momentum of either of the emitted gravitons  (labelled with indices $1$ or $2$ in \figref{Feynman-diagrams-for-annihilation})  having energy $E_{\text{GW}}=k_{\text{GW}}$, $\Delta N=2$ is the number of gravitons produced per interaction, $f_{\text{GW}}$ is the 1-particle density function for gravitons and $D_{2}$ is the 2-body phase space volume of the outgoing gravitons. 
Combining \eqsref{BTE-GWS-1} - \refb{BTE-GWS-2} leads us to the Boltzmann equation for GW from annihilation (\appnref{Ampl-GWS-spectrum})
\begin{equation}
\dot{\rho}_{\text{GW}}^{\rm ann.}+4\mathcal{H}\rho_{\text{GW}}^{\rm ann.}=\frac{\omega\left(t\right)\rho_{\phi}^{2}}{4\pi M_{p}^{4}}\Sigma^{k},\label{eq:BTEGWS}
\end{equation}
where, $\Sigma^{k}=\sum_{n}n\left|\mathcal{P}^{k}_{n}\right|^{2}$, $\mathcal{P}^{k}_{n}$ is $n$th Fourier coefficient of $\mathcal{P}^{k}$ and $\omega\left(t\right)$ is defined in \eqref{freqinflaton}.
We can solve this Boltzmann equation numerically along with \eqref{BTEinflaton} to get the energy density of gravitational waves at the end of reheating, here denoted by $\rho_{\rm GW}^{\rm ann.}\left(a_{\rm RH}\right)$, and estimate the frequency distribution $\frac{d\rho_{\rm GW}^{\rm ann.}}{d\ln{f}}$ at the end of reheating. 
The details of this procedure is given in \appnref{GWD-spectrum}. 
After the end of reheating, the spectrum of energy density redshifts as $\left(\frac{\xi T_{0}}{T_{\text{RH}}}\right)^{4}$. 
We can approximate the spectral behaviour of the density parameter of such GWs as 
\begin{equation}
\Omega_{\text{GW}}^{\rm ann.} h^{2}=\left(\frac{h^{2}}{\rho_{c,0}}\right)\left(\frac{\xi T_{0}}{T_{\text{RH}}}\right)^{4}C\left(f\right)\left.\frac{d\rho_{\text{GW}}^{\rm ann.}}{d\left(\ln{f}\right)}\right|_{\rm RH},\label{eq:omegaGWscattering}
\end{equation}
where $C\left(f\right)$ is a cut-off function that mimics the evolution of inflaton energy density after reheating ends.
We note here that the emitted gravitons are monochromatic at each point in the evolution of the Universe due to the 2-body kinematics of the final states. 
Consequently, we can define characteristic frequencies of gravitons observed today that were emitted at the end of inflation and reheating as $f_{e}=\frac{\omega\left(a_\text{end}\right)}{2\pi}\frac{a_\text{end}}{a_{\rm RH}}\frac{\xi T_{0}}{T_{\text{RH}}}$ and $f_\text{RH}=\frac{\omega\left(a_\text{end}\right)}{2\pi}\frac{\xi T_{0}}{T_{\text{RH}}}$ respectively. 
For $k=2$, the spectrum at the end of reheating is $\frac{d\rho_{\rm GW}}{d\left(\ln f\right)}=\frac{1}{2}\sqrt{\frac{f_{e}}{f}}\rho_{\rm GW}^{\rm ann.}\left(a_{\rm RH}\right)$ as derived in \eqref{spectrumRHGWS} and the cut-off function is given by $C\left(f\right) = e^{-\frac{5}{4}\sqrt{\alpha}\left(\frac{f}{f_\text{RH}}\right)^{2}}$,  with $\alpha$ given in \eqref{def-temperature}.

Finally, to get the total spectrum of the gravitational waves observed today, we can add up the contributions from decay and annihilation processes each since these are independent. 
Adding \eqsref{omegaGWdecay} and \refb{omegaGWscattering} we find the total $\Omega_{\text{GW}} h^{2}$ spectrum 
\begin{equation}
\Omega_{\text{GW}} h^{2}=\left(\frac{h^{2}}{\rho_{c}}\right)\left(\frac{\xi T_{0}}{T_{\text{RH}}}\right)^{4}\left(\left.\frac{d\rho_{\rm GW}^{\rm D}}{d\ln{f}}\right|_{\rm RH}+C\left(f\right)\left.\frac{d\rho_{\text{GW}}^{\rm ann.}}{d\left(\ln{f}\right)}\right|_{\rm RH}\right).\label{eq:omegaGWtotal}
\end{equation}

\subsection{Leptogenesis during RHN-Assisted Reheating}
\label{sec:leptogenesis}
As discussed earlier, after the end of inflation, the reheating process is governed by the decay of the inflaton to RHNs and a subsequent decay of RHNs to SM particles via RHN-Yukawa interaction in \eqref{interactionlagrangian}, making the reheating phase as an extended one. 
Note that due to the Majorana nature of the RHNs, its out-of-equilibrium decay during the reheating process can also generate a finite amount of lepton asymmetry provided the required CP-violation is present in neutrino Yukawa interaction. 
In general in the context of extended reheating, depending on the RHN masses $M_{i}$, RHNs produced from the inflaton decay would find themselves in a situation, where the condition (a) $T_{\rm max}>M_{i}>T_{\rm RH}$, or (b) $M_{i}>T_{\rm max}>T_{\rm RH}$ is satisfied, with $T_{\rm max}$ representing the maximum temperature of the Universe just after the end of inflation. In our setup, although the decay of the RHN, produced from the inflaton decay, generates the radiation bath, the effective source of radiation energy is nothing but the energy stored in the inflaton itself at the end of inflation. As a consequence, both the scenarios  
(a) and (b) are possible to be realised here. 
For situation (a), additional RHNs can also be produced via inverse decay during the period $T_{\rm max}>T>M_{i}$, which carries a possibility to bring them in equilibrium with the thermal bath.
Subsequently, around $T\lesssim M_{i}$ all these RHNs would effectively decay. 
On the other hand, for scenario (b), since all the RHNs are produced non-thermally from the inflaton decay, 
their decay starts immediately after being produced from inflaton decay and continues till $\Gamma_{N_i}=\mathcal{H}$ is satisfied. 

The decay of the RHNs in both cases produces an effective CP asymmetry which is expressed as \cite{Davidson:2008bu}
\begin{equation}
\epsilon_{\ell}^{i}=\frac{\Gamma(N_{i}\to\ell_{L}+H)-\Gamma(N_{i}\to\bar{\ell}_{L}+H^{\dagger})}{\Gamma(N_{i}\to\ell_{L}+H)+\Gamma(N_{i}\to\bar{\ell}_{L}+H^{\dagger})}=\frac{1}{8\pi K_{ii}}\sum_{j\neq i}{\rm Im}[K_{ij}^{2}]\mathcal{F}\left(\frac{M_{j}^{2}}{M_{i}^{2}}\right),\label{eq:epsilonCP}
\end{equation}
where $K=(Y_{\nu}^{\dagger}Y_{\nu})$ and $\mathcal{F}(x)=\sqrt{x}\left[1+\frac{1}{1-x}+(1+x)\ln\left(\frac{x}{1+x}\right)\right]$ represents the loop function generated due to the interference between the decay process of $N_{i}$ at tree level and one loop level considering both the vertex and the self energy corrections for hierarchical RHNs.
Note that, for scenario (a), decay of all generations of RHNs may not in general contribute to the total CP asymmetry.  For example, within the framework of hierarchical RHN masses following the condition $M_{1}<M_{2}<\ldots<M_{i}$, the produced CP asymmetry from $N_{i>1}$ decay may get diluted due to the still prevailing production process of $N_{1}$ (via inverse decay). 
In that case, only the lightest RHN would contribute for non-zero CP asymmetry. 
On the contrary, with $M_{i}>T_{\rm max}$ satisfied in scenario (b),  all RHNs effectively generate the CP asymmetry, and total CP asymmetry would be dictated by $\epsilon_{\ell}=\sum_{i}\epsilon_{\ell}^{i}$ with $i=1,\,2,\,3,\ldots$. Remember that, in our setup, we have only two RHNs, out of which inflaton can decay to only lightest RHN as $M_2>m_\phi>M_1$. Consequently, CP asymmetry can only be generated from the lightest RHN for both the scenarios.

The CP asymmetry generated in this way leads to an effective leptonic (more specifically $B-L$) asymmetry which eventually gets converted to baryon asymmetry due to sphaleron transitions at the time of sphaleron decoupling temperature $T_\text{sp}\sim131.7$ GeV~\cite{DOnofrio_2014}. 
Since, in our case, the generated CP asymmetry (and eventually the $B-L$ asymmetry) from the RHN decay are intricately connected with the decay of the inflaton and the Hubble expansion rate during reheating process, solution of the coupled Boltzmann equations related to the energy density of the inflaton $\rho_{\phi}$, number density of the RHNs $n_{N_{i}}$ and energy density of the radiation $\rho_{R}$ (as presented in \eqsref{phi}-\refb{r}) is required along with the BE for $B-L$ asymmetry which reads as
\begin{align}
	\frac{dn_{B-L}}{dt}+3\mathcal{H}n_{B-L} & =-\sum_{i}\Gamma_{N_{i}}\left[\epsilon_{\ell}^{i}(n_{N_{i}}-n_{N_{i}}^{{\rm eq}})+\frac{n_{N_{i}}^{{\rm eq}}}{2n_{\ell}^{{\rm eq}}}n_{B-L}\right],
    \label{eq:b-l}
\end{align}
to estimate the exact $B-L$ asymmetry surviving till $T_\text{sp}$.
Finally, the resultant baryon asymmetry is evaluated as \cite{Harvey_1990} $n_B^{\rm sp}=(28/79) n_{B-L}^{\rm sp}$.

\section{Gravitational Wave Spectrum and Leptogenesis}\label{sec:sec4}

\subsection{Phenomenology of Leptogenesis during Reheating}
\label{sec:numeric-leptogenesis}

As discussed earlier, the parameters $k$, $y_{\phi},~M_{1}$ and $Y_{\nu}$ dictate the reheating dynamics in our setup, where the structure of neutrino Yukawa coupling $Y_\nu$ depends on the number of RHNs present in the system (for type-I seesaw framework). Since we consider the presence of two hierarchical RHNs in our system, the corresponding Yukawa coupling matrix $Y_{\nu}$ can be evaluated using the well known CI parametrisation \cite{Casas_2001} as
\begin{equation}
Y_{\nu}=-i\frac{\sqrt{2}}{v}UD_{\sqrt{m}}R^{T}D_{\sqrt{M}},\label{eq:CI-1}
\end{equation}
where $U$ is the Pontecorvo-Maki-Nakagawa-Sakata (PMNS) matrix connecting the flavour basis with the mass basis of light neutrinos. Here, $D_{\sqrt{m}}=\diag{[0,\sqrt{m_2},\sqrt{m_3}]}$ and $D_{\sqrt{M}}=\diag{\left[\sqrt{M_{1}},\sqrt{M_{2}}\right]}$ represent the diagonal matrices containing the square root of light and heavy neutrino mass eigenvalues respectively setting $m_1 = 0$ for normal hierarchy. The two other light neutrino masses ($m_{2,3}$) are to be obtained using the data on solar and atmospheric mass-squared differences ($\Delta m_{21}^2$ and $\Delta m_{31}^2$ respectively) while for heavy RHNs, $M_1$ is considered as a parameter with their ratio ($M_2/M_1$) kept as $10^5$, throughout the analysis. 
Within this minimal scenario, the orthogonal matrix $R$ is parametrised using one complex angle $\theta=\theta_{r}+{\it i}\theta_{i}$ by
\begin{equation}
    R=\left(\begin{array}{ccc}
0&\cos{\theta} & \sin{\theta}  \\
0 & -\sin{\theta} & \cos{\theta}  
\end{array}\right).
\end{equation}
To evaluate the structure of $Y_\nu$, the best fit values of mixing angles in $U$ and two mass-squared differences corresponding to normal hierarchical scenario are used from neutrino oscillation data \cite{Esteban_2024}.

Thus we have effectively five free parameters in our study, namely $k$, $M_{1}$, $y_{\phi}$, $\theta_{r}$, and $\theta_{i}$.
To start with, we fix $k=2$, which along with $N_*=55$ leads to $\lambda= 2\times 10^{-11}$ as well as a constant inflaton mass (see footnote~\ref{Footnote:lambda-calc} and \eqref{massinflaton}). 
Moreover, the Yukawa coupling $y_{\phi}$ is kept fixed at the largest possible value $y_{\phi}=10^{-6}$ so as to prevent preheating~\cite{Greene_1997, Greene_1998, Dufaux_2006, Drewes_2017, Garcia_2021} and still give us gravitational wave signal in the detectable range. 
Lowering the $y_{\phi}$ results in smaller $T_{\rm RH}$ value and the spectrum shifts to higher frequency range which is beyond the reach of current and future detectors. 
Here, the allowed range of $M_{1}$ turns out to be between $10^{13}$ GeV and $10^{9}$ GeV. 
Any heavier RHN will not allow the inflaton (of mass $\sim 10^{13}$ GeV) to decay. 
On the other hand, RHNs lighter than $M_1= 10^9$ GeV cannot be responsible for successful leptogenesis. 
Within this range of masses of the RHNs, we have scanned and extracted values of the CI parameters for which we get the observed baryon asymmetry of the Universe. 
We find that $T_{\rm RH}$ is practically insensitive to the CI parameters unless there is a RHN domination which happens for $M_{1}$ much lower than ${10}^{9}\text{ GeV}$ only. 
Here we mention in passing that the GW spectra is found to be dependent on these free parameters only through $T_{\rm RH}$, and as a consequence, is also insensitive to the CI parameters to some extent, within the allowed range of $M_1$.

\begin{table}[h]
\centering{}%
\begin{tabular}{|c|c|c|c|}
\hline 
Benchmark Point (BP) & $M_{1}\text{ (GeV)}$  & $\theta$  & $T_{\text{RH}}\text{ (GeV)}$\tabularnewline
\hline 
BP1 & $10^{11}$  & $2.83+0.0035i$  & $1.79\times10^{8}$\tabularnewline
\hline 
BP2 &$10^{10}$  & $2.83+0.006i$  & $9.951\times10^{7}$\tabularnewline
\hline 
BP3 & $10^{9}$  & $2.83+0.012i$  & $5.73\times10^{7}$\tabularnewline
\hline 
\end{tabular}\caption{Our chosen benchmark points satisfying the correct baryon asymmetry of the Universe observed today.}\label{tab:benchmarkpoints}
\end{table}
In \tabref{benchmarkpoints}, we include three benchmark points (BP) for $M_{1}$ and corresponding $\theta$ parameters (with $\theta_{r}$ and $\theta_{i}$ values) involved in the $R$-matrix for which leptogenesis during reheating produces correct baryon asymmetry of the Universe observed today.
After solving the relevant BEs for these BPs, we proceed to analyse the variation of energy density of different components of the Universe \emph{w.r.t.} the normalised scale factor. Such variations for BP1 and BP2 are shown in the left and right panels of \figref{Evolution-of-energy-densities} respectively. Both plots indicate that the energy density of the inflaton field ($\rho_\phi$ is represented by blue curve) initially undergoes a slow decreasing phase, followed by a rapid fall 
in its energy density when the decay rate of the inflaton becomes comparable to the Hubble expansion rate of the Universe \emph{i.e.}, $\frac{\Gamma_{\phi}}{\mathcal{H}}\gtrsim 1$. 

Interestingly, during the early stage of the inflaton oscillation (\emph{i.e.}, for small values of $\mathcal{A}=\frac{a}{a_{\rm end}}$), the production of the most energetic RHNs takes place. 
As a result, the energy density of these RHNs ($\rho_{N_1}$, marked by the red curve) reaches its peak just after the end of inflation the immediate decay of which is responsible for creating a peak in the $\rho_R$ as well, characteristic of maximum temperature of the Universe $T_{\rm max}$ after inflation. It is interesting to note that these RHNs  acquire large field dependent mass correction $M_1(\phi)$ on top their bare mass $M_1$ as in \eqref{massRHN} due to the presence of large 
$\rho_\phi$ at smaller $\mathcal{A}$. This enhancement in RHN mass leads to a significant increase of decay rate of the RHNs $\Gamma_{N_1}$ initially. Subsequently, with the decreasing $\rho_{\phi}$, which affects the $M_1(\phi)$, we observe a sizeable decline in the $\rho_{N_1}$ as evidenced in both plots. The energy density of radiation ($\rho_R$, denoted by green curve) thus decreases relatively fast (compared to usual decrease due to Hubble expansion) from the end of inflation till $\mathcal{A}^*$, where  $\mathcal{A}^*$ (as indicated by the vertical dashed line) is defined as the scale factor where $M_1(\phi)$ becomes subdominant compared to the RHN bare mass $M_1$ alone\footnote{In this work, the numerical value of $\mathcal{A}^*$ is precisely extracted from the condition $M_1(\phi)= M_1/10 $.}.
Notice that the value of $\mathcal{A}^*$ for $M_1=10^{11}$ GeV is smaller than that for $M_1=10^{9}$ GeV. This is due to the fact that for smaller $M_1$, field dependent contribution  $M_1(\phi)$ remains dominant for a longer time compared to the case for larger $M_1$. 

\begin{figure}[t]
\includegraphics[width=0.5\textwidth]{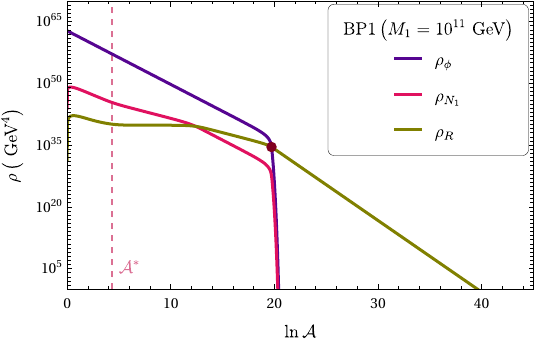}\includegraphics[width=0.5\textwidth]{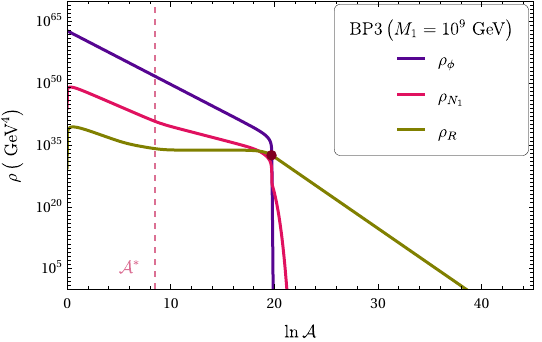}
\caption{Evolution of energy densities of different components \emph{w.r.t.} renormalised scale factor $\mathcal{A}=a/a_{{\rm end}}$ for BP1 (left panel) and BP3) (right panel) respectively.} 
\label{fig:Evolution-of-energy-densities}
\end{figure}

As time progresses, the field dependent mass of the RHN decreases and beyond the threshold $\mathcal{A}=\mathcal{A}^*$, the bare mass of the RHN starts to influence. Consequently, $\Gamma_{N_1}$ becomes smaller than the situation when $M_1(\phi)$ dominated over $M_1$. This reduction in decay rate results in a slow decrease of the $\rho_{N_1}$ compared to the $\mathcal{A}<\mathcal{A}^*$ phase, which is also evident from the change in the slope of $\rho_R$  around $\mathcal{A}=\mathcal{A}^*$ (marked by the vertical red dashed lines) in both panels. At a later stage when the Hubble expansion rate becomes comparable to $\Gamma_{N_1}$ (which is dominantly influenced by $M_1$), RHNs decay rapidly and its energy density dilutes completely around $\mathcal{A}\sim10^{20}$. Notice that for both BP1 and BP3, RHNs  decay away completely while $\rho_\phi$ remains dominant over other energy densities. Thus, the onset of the radiation domination in both BPs can be determined solely from the condition $\rho_\phi= \rho_R$.  In fact, within our framework, this conclusion holds for RHN masses $M_1> 10^9$ GeV as no RHN domination was observed for these RHN masses.

\begin{figure}[h]
\includegraphics[width=0.5\textwidth]{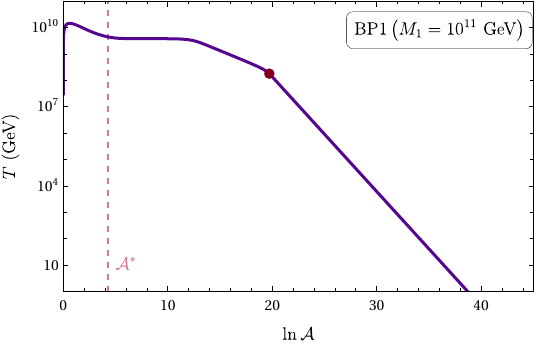}\includegraphics[width=0.5\textwidth]{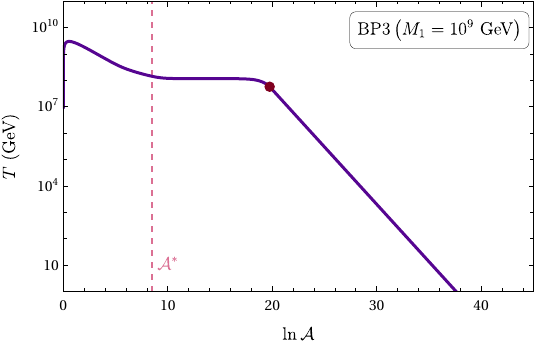}\caption{Evolution of Temperature of the Universe for BP1 ($M_{1}=10^{11}\text{ GeV}$) and BP3 ($M_{1}=10^{9}\text{ GeV}$) (from left to right) respectively.}\label{fig:Evolution-of-Temperature}
\end{figure}

We can also estimate the evolution of temperature of the Universe following \eqref{def-temperature} using the evolution of the energy density of the SM plasma $\rho_{R}$ obtained above, which is shown in \figref{Evolution-of-Temperature} for BP1 (left panel) and BP3 (right panel). 
Notice that for both the BPs, after the end of inflation, the temperature rises quickly to a maximum value $T_{\rm max}$ due to the production of the standard model particles (from $N$ mediated decay of $\phi$) and their instantaneous thermalisation. The subsequent variation of the temperature reflects the corresponding change in $\rho_{R}$ since $T\propto\rho_{R}^{\frac{1}{4}}$. As the Universe continues to evolve, radiation energy starts to dominate over $\rho_\phi$. The temperature which indicates the commencement of such radiation domination is known as reheating temperature $T_{\rm RH}$ and is determined from the condition $\rho_\phi= \rho_R$ in all the BPs (marked by `$\bullet$' in \figref{Evolution-of-energy-densities} and \figref{Evolution-of-Temperature}).
Furthermore, as discussed earlier, for lighter $M_{1}(t)$, $\Gamma_{N_1}$ decreases. and as a result, the completion of reheating process is delayed. This induces a lower reheating temperature as can be seen in \tabref{benchmarkpoints} as well as in the \figref{Evolution-of-Temperature}.
Subsequently, once the reheating is complete and the energy density of the thermal bath starts to dominate, the temperature of the bath evolves as $T\propto \mathcal{A}^{-1}$.
This relation is useful in determining the amount of redshift $\frac{a_{\rm RH}}{a_{0}}=\frac{\xi T_{0}}{T_{\rm RH}}$ encountered post-reheating, where $\xi$ describes the change in the number of species in the thermal bath in this period. 

\begin{figure}[b]
\includegraphics[width=0.5\textwidth]{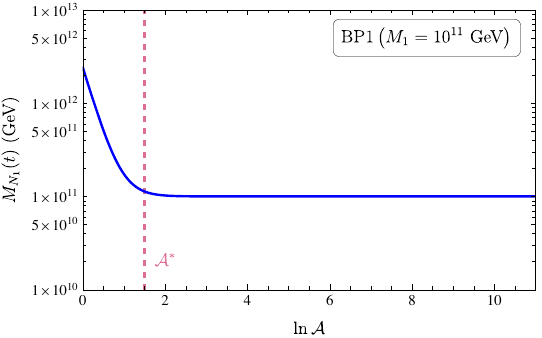}
\includegraphics[width=0.5\textwidth]{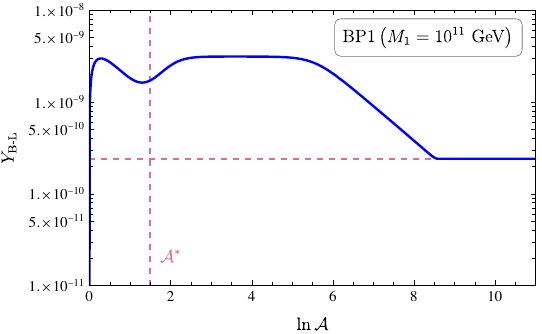}
\caption{Evolution of  $M_{N_1}(t)$ (left panel) and $Y_{B-L}=n_{B-L}/s$ (right panel) for BP1($M_{1}=10^{11}\text{ GeV}$).}\label{fig:Evolution-of-nBL}
\end{figure}

Finally, in left panel of \figref{Evolution-of-nBL}, we plot the variations of the total RHN mass $M_{N_1}(t)$ which would be helpful in explaining the evolution of $Y_{B-L}=n_{B-L}/s$ as a function of the scale factor for BP1, where $s$ is the entropy density of the Universe. This typical variation of $Y_{B-L}$ is a characteristic of our RHN-assisted reheating scenario. As previously discussed, due to the presence of large inflaton energy density, RHN gains dominant field dependent mass $M_1\left(\phi\right)$ leading to a large $M_{N_1}(t)$ initially just after the end of inflation, as evident from the left panel of \figref{Evolution-of-nBL}. As a consequence, decay of the RHNs produced during this initial period  generates a large amount of CP asymmetry, resulting an initial peak in the $B-L$ asymmetry as shown in the right panel of \figref{Evolution-of-nBL}. The subsequent decrease in $M_{N_1}\left(t\right)$ induces a fall in $Y_{B-L}$
till $\mathcal{A^*}$, beyond which it increases again and saturates for some time (represented by the plateau region). This portion of the increment is related to the increase in co-moving number density of the RHNs at this stage from the decay of inflaton. As the Universe evolves further, a stage is reached when $B-L$ asymmetry generation rate becomes comparable to the Hubble expansion rate making it effectively constant.

A further dilution of $Y_{B-L}$ occurs at a later stage of reheating when all the RHNs effectively decayed away injecting entropy to the SM thermal bath. The final $B-L$ asymmetry results after the end of reheating which gets converted to the observed baryon asymmetry of the Universe (indicated by horizontal dashed magenta line in right plot of \figref{Evolution-of-nBL}). Note that since for BP1, $T_{\rm max}<M_{1}$  holds, no dilution of asymmetry occurs due to the absence of inverse decay. This observation holds for other two BPs as well. Furthermore, reducing the bare mass of the RHN below $10^9$ GeV can not only lead to the RHN dominated phase following the inflaton domination, but also the condition $T_{\rm max}>M_{1}>T_{\rm RH}$. In that case, inverse decay can further contribute to the dilution of the asymmetry produced from decay. Additionally, the RHN domination prior to their decay may result in significant entropy injection, which can also play important role in diluting the $B-L$ asymmetry during the RHN decay. As a result, it is found to be challenging to generate the observed baryon asymmetry from the decay of RHN having $M_1<10^9$ GeV.

\subsection{Combined GW Spectrum for $k=2$}

We now proceed to discuss the GW spectra obtained from the decay and annihilation of the inflaton field during reheating while taking into the consideration the simultaneous satisfaction of the BAU via leptogenesis from the decay of RHNs which is intricately connected to the reheating dynamics in our set-up. 
The GW spectrum resulting solely from the inflaton decay is depicted by the dashed lines in the left panel of \figref{GW-frequency-spectrum-k-2}, corresponding to the three BPs (for RHN mass and other parameters) of \tabref{benchmarkpoints}. 
On the other hand, the dotted lines describe the spectrum obtained from inflaton annihilation only in $\Omega_{\rm GW} h^2 - f$ plane, with $f$ indicating the frequency of the GW as of today. 
As discussed earlier (following \eqref{f0}), such GW spectrum (originated from both decay and annihilation) is obtained after taking care of redshift from its production point with frequency $f\left(a\right)$ till $T_{\rm RH}$ in first phase (during extended period of reheating) and thereafter, including the redshift in the radiation dominated era until today. 

\begin{figure}[t]
\centering
\includegraphics[width=0.49\textwidth]{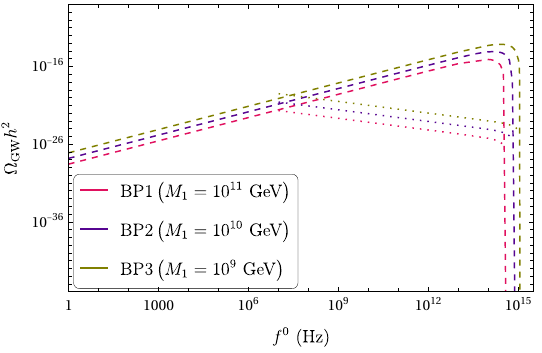}
\includegraphics[width=0.49\textwidth]{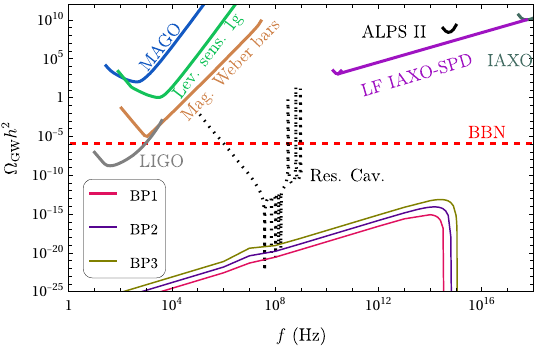}
\caption{GW frequency spectrum for BP1, BP2 and BP3 given in \tabref{benchmarkpoints}. 
The dotted and dashed lines on the left panel are due to inflaton annihilation and decay processes respectively, while the solid lines of the same colours in the right panel are for their combined spectrum.
The sensitivity bounds for the future GW experiments \cite{Arvanitaki:2012cn,Goryachev:2014yra,Aggarwal:2020olq,Aggarwal:2020umq,Domcke:2020yzq,Ringwald:2020ist,Berlin:2021txa,Domcke:2022rgu,Herman:2022fau,Aggarwal_2025} are indicated by solid lines while a black dotted line (being overly optimistic as criticized in the literature~\cite{Aggarwal_2025}) is used to indicate the projected sensitivity of the resonant cavity experiment~\cite{Herman_2023}.} \label{fig:GW-frequency-spectrum-k-2}.
\end{figure}

As can be seen from the dashed line in the left panel of \figref{GW-frequency-spectrum-k-2}, the decay spectrum increases with a power law ($\propto f$) and then decreases sharply after reaching a peak value, for a fixed RHN mass. 
Gravitons with frequencies larger than this peak frequency are never produced due to the insignificant inflaton number density at the end of reheating, resulting such a sharp fall in amplitude beyond this peak frequency. On the other hand, we impose a lower frequency cut-off $\sim 1$ Hz while plotting the GW spectrum in Fig.~\ref{fig:GW-frequency-spectrum-k-2}, which can be understood in the following way. Note that the rate of the $1 \rightarrow 3$ inflaton decay process ($\Gamma^{1\to3}$) becomes divergent due to the contribution from gravitons with energy approaching zero.
Traditionally, this IR divergence goes away when loop correction involving virtual gravitons to two body decay of inflaton (producing two RHNs only) is taken into account~\cite{Weinberg:1965nx}.
However, this is inconsequential to our analysis since the relevant quantity in the RHS of \eqref{BTEGWD} is the power spectrum $E_k \frac{d\Gamma^{1\rightarrow 3}}{dE_k}$ only, which remains finite.
Additionally, gravitons become super-horizon above a certain wavelength given by the Hubble scale $\mathcal{H}^{-1}$. Since the super-horizon modes of the gravitational waves freeze out (\emph{i.e.}, become constant)~\cite{Domcke:2024soc} in the expanding Universe, they may be excluded.
This gives us a mere lower frequency cut-off (for representation purpose in Fig. \ref{fig:GW-frequency-spectrum-k-2} only) at $\frac{a_\text{RH}}{a_0}\frac{\mathcal{H}_\text{RH}}{2\pi}\approx1~\text{Hz}$ using the estimate of the Hubble scale at the time of reheating~\cite{Huang:2019lgd,Tokareva:2023mrt} with $T_\text{ RH}\sim 10^8$ GeV.

Turning our attention to the spectrum obtained from inflaton annihilation (dotted lines of Fig.~\ref{fig:GW-frequency-spectrum-k-2}, left panel), we note that there exists a low frequency cut-off near $f_{e}=10^{7} \text{ Hz}$ in this case as well which characterises the  beginning of inflaton oscillation before which no gravitons are produced. The spectra then decrease with the power law ($ \propto 1/\sqrt{f}$) as shown in \appnref{GWD-spectrum} and then falls sharply 
due the exponential decrease of the inflaton density at the end of reheating (similar to the GW spectrum from inflaton decay). Note that such annihilation spectra of GWs is the artifact of the extended reheating scenario of our framework. 
In case of instantaneous reheating, it would emit GW of unique frequency only, characteristic of inflaton mass. Contrary to this, in case of extended reheating, GWs emitted at different points during reheating period would redshift differently and govern the spectrum. 

Now let us discuss the behaviour of the spectrum for different BPs. As indicated by \eqref{omegaGWdecay} and \eqref{omegaGWscattering}, current energy density of the GW produced from inflaton decay and annihilation during reheating depends on  the reheating temperature as $\Omega_{\text{GW}}h^{2}\propto T_{\text{RH}}^{-4}$.
Moreover, it has been already observed that $T_{\rm RH}$ in our set-up primarily depends on the mass of the lightest RHN $M_{1}$.
More specifically, A smaller value of $M_{1}$ leads to a lower value of $T_{\text{RH}}$. Thus, decreasing $M_{1}$ increases the density parameter $\Omega_{\text{GW}}h^{2}$.
Furthermore, with lower reheating temperature, frequency of the GW redshifts less compared to scenarios with a higher $T_{\rm RH}$ thereby producing a gravitational wave spectrum characterised by higher current frequencies as observed in \figref{GW-frequency-spectrum-k-2}.

Finally, the small bump around $f \sim10^{7}-10^{8}\text{ Hz}$ in the combined gravitational wave spectrum (marked as solid curves) in right panel of \figref{GW-frequency-spectrum-k-2} indicates that the contribution from inflaton annihilation becomes larger than that from inflaton decay. This frequency $f_{e}$ is characteristic to the frequency of the gravitons emitted at the end of inflation and the beginning of reheating. The reason for the larger contribution from annihilation channel is that $\rho_{\phi}$ dominates during the earlier period of reheating as expected, increasing graviton production through \eqref{BTEGWS}. 
This bump happens to fall in the detectable frequency range of the proposed resonant cavity experiment~\cite{Herman_2023}, although recent analysis~\cite{Aggarwal_2025} indicates that an appropriate comparison between the signal and the background noise may alter this sensitivity region significantly. 
Alongside, there exist other more realistic proposals~\cite{Arvanitaki:2012cn,Goryachev:2014yra,Aggarwal:2020olq,Aggarwal:2020umq,Domcke:2020yzq,Ringwald:2020ist,Berlin:2021txa,Domcke:2022rgu,Herman:2022fau,Capdevilla:2025ivz} which can probe the GW spectrum
within the frequency range specified in Fig. \ref{fig:GW-frequency-spectrum-k-2}. 
Though the $\Omega_{\text{GW}}h^2$ obtained in our analysis falls below the reach of present sensitivities of these experiments, future experiments with higher sensitivities may probe it.}

\subsubsection{Case with $k=4$}

It is important to note that for inflaton potentials characterised by $k\neq2$, the mass of the inflaton field becomes a function of its energy density, which constrains the evolution of the inflaton energy density and the subsequent RHN-assisted reheating process.
For example, in the case of $k=4$ potential, even with a sizeable inflaton-RHN coupling, the decay rate of the inflaton vanishes whenever the mass of the inflaton reduces below the mass of the RHN (denoted as $a_{*}$). 
Subsequently, the decay process of the inflaton stops, and the energy density of the inflaton starts to scale as $\rho_{\phi}\propto a^{-4}$.
In that case, the RHNs produced during the evolution of the Universe from $a=a_{{\rm end}}$ to $a=a_{*}$ cannot generate a sufficient amount of radiation whose energy density can dominate over the energy density of the inflaton. 
After being produced from the RHN decay, the energy density of the radiation then scales similarly to that of inflaton, \emph{i.e.}, $\rho_{R}\propto a^{-4}$. 
Consequently, the domination of the inflaton energy density prevails throughout the evolution of the Universe and leads to unsuccessful reheating.
For $k=6$ and above, although the field dependent mass of the inflaton does decrease below the lightest RHN mass early on, preventing the inflaton from decaying into RHNs, reheating eventually completes at much lower temperatures because of Hubble expansion.

\section{Conclusion}\label{sec:sec5}

In this study, we have provided a SM gauge invariant picture of the post-inflationary perturbative reheating process involving the neutrino sector, consisting of RHNs, responsible for generating the light neutrino masses via type-I seesaw mechanism.
In particular, reheating is driven by the decay of right-handed neutrinos (RHNs), which are produced through the decay of the inflaton field. In this context, an intriguing probe of RHN-assisted reheating in the early Universe is the potential detection of gravitational waves produced by both three-body inflaton decays and 2 $\rightarrow$ 2 annihilation processes involving inflatons during the reheating phase via bremsstrahlung. Additionally, since the later out-of-equilibrium decay of RHNs generates the thermal bath, it is natural to ask whether the same can contribute to the baryon asymmetry of the Universe via leptogen)esis. In fact, our findings indicate that the RHNs not only play a crucial role in reheating the Universe but also become effective in generating the appropriate amount of baryon asymmetry through leptogenesis, due to their out-of-equilibrium decay during the reheating epoch. Hence, linking the reheating phase to the neutrino sector, along with its potential probing through a distinctive gravitational wave spectrum and the simultaneous realisation of leptogenesis during this extended reheating period, makes the scenario particularly compelling.

The detailed analysis of such reheating reveals some interesting features associated with this scenario. In general, such an extended reheating prescription mostly depends on the shape of the inflaton potential near its minimum, whereas the shape away from the minimum determines the inflationary dynamics. 
Specific to our scenario, in addition, we observe that the non-zero RHN mass and inflaton-RHN coupling have the potential to alter the dynamics of reheating. 
For example, with $k=4$ inflaton potential near minimum, no successful reheating can be achieved due to the reduction of the effective inflaton mass (field dependent) below the RHN mass after some e-folds which is a unique observation in the context of such extended perturbative reheating. We therefore focus on the $k=2$ case in this work. 
This dominant field-dependent mass of the RHN just after the end of the inflation has played an important role in determining the initial maximum temperature of the Universe, which is found to be slightly different than in the case of zero temperature mass of the RHN only.
Furthermore, we notice that fixing the inflaton-RHN coupling to its maximum allowed value (restricted by preheating constraints) sets the reheating dynamics as practically independent of the RHN Yukawa coupling.

The simultaneous fulfilment of neutrino mass and mixing angle constraints, along with the observed baryon asymmetry of the Universe, in this case, led to a specific allowed range for the mass of the RHNs, with which we further analyse the gravitational wave spectrum. We observed that the gravitational wave spectrum generated from the decay of inflaton mostly dominates over the scattering processes. However, there exists a small region where domination of scattering contribution over the decay can be seen, indicated by the presence of small bumps in the GW spectrum (in the MHz range)  carrying a unique feature of our scenario. This region falls in the projected sensitivity range (although overly optimistic~\cite{Aggarwal_2025}) of the resonant cavity experiment~\cite{Herman_2023}.  Furthermore, the peak of the GW spectrum in this setup lies in the optical frequency range. However, the associated amplitude remains beyond the reach of other more realistic proposals~\cite{Arvanitaki:2012cn,Goryachev:2014yra,Aggarwal:2020olq,Aggarwal:2020umq,Domcke:2020yzq,Ringwald:2020ist,Berlin:2021txa,Domcke:2022rgu,Herman:2022fau,Aggarwal_2025,Capdevilla:2025ivz}. We expect the future proposals for high-frequency GW detectors may test or falsify such unique reheating scenario. Finally, we would like to comment that there exists another source of GW production via bremsstrahlung from the three body decay of RHNs \cite{Datta_2024, Murayama_2025}
 while generating the thermal bath and contributing toward leptogenesis. However, such a contribution is found to be subdominant compared to the GW spectrum contributed by three body decay of inflaton. 

\section{Acknowledgement}
The work of AD is supported in part by Basic Science Research Program through the National Research Foundation of Korea (NRF) funded by the Ministry of Education, Science and Technology (NRF-2022R1A2C2003567). 
The work of SK is partially supported by Science, Technology $\&$ Innovation Funding Authority (STDF) under grant number 48173.
AS acknowledges the support of the Institut Henri Poincaré (UAR 839 CNRS-Sorbonne Université), and LabEx CARMIN (ANR-10-LABX-59-01). AS also acknowledges the hospitality of CERN theory division during a visit where part of this work was done. The work of AS is supported during its initial stage by the grants CRG/2021/005080 and MTR/2021/000774 from SERB, Govt. of India.

\appendix

\section{Feynman Rules Involving Gravitons}
\label{sec:Feynman-rules}

The action functional for all fields are given in \eqref{actionfunctional-full}.
We treat the metric $g_{\mu\nu}$ as the dynamical field describing gravity and perturb it \emph{w.r.t.} a background metric $g_{\mu\nu}=\eta_{\mu\nu}+\kappa h_{\mu\nu}$.
The second term in \eqref{actionfunctional-full}, $\sqrt{-g}\mathcal{R}$, upon expansion generates the graviton propagator and the triple graviton vertex as:
\begin{figure}[H]
\begin{minipage}[c]{0.3\linewidth}
\includegraphics{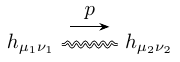}
\end{minipage}
\begin{minipage}[c]{0.69\linewidth}
\begin{equation}
D_{\mu_{1}\nu_{1}\mu_{2}\nu_{2}}=\frac{i}{2p^{2}}\left(\eta_{\mu_{1}\mu_{2}}\eta_{\nu_{1}\nu_{2}}+\eta_{\mu_{1}\nu_{2}}\eta_{\mu_{2}\nu_{1}}-\eta_{\mu_{1}\nu_{1}}\eta_{\mu_{2}\nu_{2}}\right)
\end{equation}
\end{minipage}

\begin{minipage}[c]{0.3\linewidth}
\includegraphics{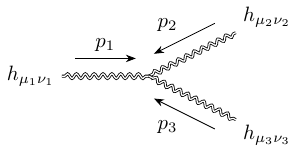}
\end{minipage}
\begin{minipage}[c]{0.69\linewidth}
\begin{equation}
\begin{split}i\kappa\left(\text{Sym}\right)\left[-\frac{1}{2}P_{6}\left(p_{1}\cdot p_{2}\eta_{\mu_{1}\nu_{1}}\eta_{\mu_{2}\mu_{3}}\eta_{\nu_{2}\nu_{3}}\right)\right.\\
-\frac{1}{2}P_{3}\left(p_{1}\cdot p_{2}\eta_{\mu_{1}\mu_{2}}\eta_{\nu_{1}\nu_{2}}\eta_{\mu_{3}\nu_{3}}\right)\\
+\frac{1}{4}P_{3}\left(p_{1}\cdot p_{2}\eta_{\mu_{1}\nu_{1}}\eta_{\mu_{2}\nu_{2}}\eta_{\mu_{3}\nu_{3}}\right)\\
+2P_{3}\left(p_{1}\cdot p_{2}\eta_{\mu_{1}\nu_{3}}\eta_{\mu_{2}\nu_{1}}\eta_{\mu_{3}\nu_{2}}\right)\\
P_{3}\left(p_{1\mu_{3}}p_{2\nu_{3}}\eta_{\mu_{1}\mu_{2}}\eta_{\nu_{1}\nu_{2}}\right)\\
\left.-P_{6}\left(p_{1\mu_{3}}p_{2\mu_{1}}\eta_{\mu_{2}\nu_{1}}\eta_{\nu_{2}\nu_{3}}\right)\right]
\end{split}
\end{equation}
\end{minipage}
\end{figure}
Expanding the matter action \eqref{actionfunctional-full} in the same way, we obtain the following inflaton-graviton and RHN-graviton couplings giving rise to GWs. 
\footnote{Here $I_{\mu\nu\alpha\beta}=\frac{1}{2}\left(\eta_{\mu\alpha}\eta_{\nu\beta}+\eta_{\mu\beta}\eta_{\nu\alpha}\right)$ is a symmetrisation tensor.}
\begin{figure}[H]
\begin{minipage}[c]{0.3\linewidth}
\includegraphics{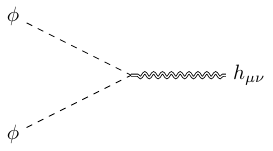}
\end{minipage}
\begin{minipage}[c]{0.69\linewidth}
\begin{equation}
T_{\mu\nu}\left(\phi\right)=-i\kappa\left(\partial_{\mu}\phi\partial_{\nu}\phi-\eta_{\mu\nu}\left(\frac{1}{2}\partial^{\mu}\phi\partial_{\mu}\phi-V\left(\phi\right)\right)\right)\label{eq:feynrulesphigrav1}
\end{equation}
\end{minipage}

\begin{minipage}[c]{0.3\linewidth}
\includegraphics{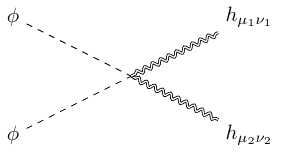}
\end{minipage}
\begin{minipage}[c]{0.69\linewidth}
\begin{equation}
\begin{split}V_{\mu_{1}\nu_{1}\mu_{2}\nu_{2}}\left(\phi\right)=i\kappa^{2}\left[2I_{\mu_{1}\nu_{1}\rho\xi}I_{\sigma\mu_{2}\nu_{2}}^{\xi}\partial^{\rho}\phi\partial^{\sigma}\phi\right.\\
-\frac{1}{2}\left(\eta_{\mu_{1}\nu_{1}}I_{\rho\sigma\mu_{2}\nu_{2}}+\eta_{\mu_{2}\nu_{2}}I_{\rho\sigma\mu_{1}\nu_{1}}\right)\partial^{\rho}\phi\partial^{\sigma}\phi\\
\left.-\frac{1}{2}\left(I_{\mu_{1}\nu_{1}\mu_{2}\nu_{2}}-\frac{1}{2}\eta_{\mu_{1}\nu_{1}}\eta_{\mu_{2}\nu_{2}}\right)\left(\partial^{\mu}\phi\partial_{\mu}\phi-V\left(\phi\right)\right)\right]
\end{split}
\label{eq:feynrulesphigrav2}
\end{equation}
\end{minipage}
\end{figure}

\begin{figure}[H]
\begin{minipage}[c]{0.3\linewidth}
\includegraphics{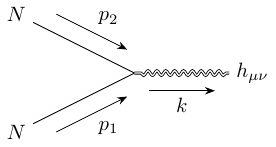}
\end{minipage}
\begin{minipage}[c]{0.69\linewidth}
\begin{equation}
\frac{i\kappa}{8}\left(2\eta_{\mu\nu}\left(\slashed{p_{1}}+\slashed{p_{2}}-2m\right)-\left(p_{1}+p_{2}\right)_{\mu}\gamma_{\nu}-\gamma_{\mu}\left(p_{1}+p_{2}\right)_{\nu}\right)
\end{equation}
\end{minipage}

\begin{minipage}[c]{0.3\linewidth}
\includegraphics{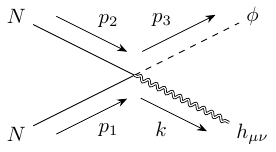}
\end{minipage}
\begin{minipage}[c]{0.69\linewidth}
\begin{equation}
i\kappa y_{\phi}\eta_{\mu\nu}\phi\label{eq:feynruledecay2}
\end{equation}
\end{minipage}
\end{figure}

\section{The $\phi\rightarrow NNh$ Decay Processes}
\label{sec:Ampl-GWD-spectrum}
In this section we will discuss how different Feynman diagrams in \figref{Feynman-diagram-for-decay} contribute to the total differential decay width in \eqref{diffdecaywidth}. To start with, let us discuss the fate of  the last two diagrams in \figref{Feynman-diagram-for-decay}. 
Using the Feynman rule in \eqref{feynruledecay2} and the traceless-ness condition in \eqref{conditiontraceless}, we get the Feynman amplitude for the $3$rd Feynman diagram as
\begin{align}
i\mathcal{M}_{3}=y_{\phi}\eta_{\mu\nu}\epsilon^{\mu\nu}\bar{u}\left(p_{3}\right)v\left(p_{4}\right)=0.
\end{align}
For the last Feynman diagram, the Feynman amplitude is given by
\begin{align}
i\mathcal{M}_{4}=y_{\phi}T_{\mu\nu}\left(\phi\right)\epsilon^{\mu\nu}\frac{i}{\left(p_{1}-k\right)^{2}}\bar{u}\left(p_{3}\right)v\left(p_{4}\right).
\end{align}
Since,
\begin{align}
T_{\mu\nu}\epsilon^{\mu\nu} & =-i\kappa\left(\partial_{\mu}\phi\partial_{\nu}\phi\epsilon^{\mu\nu}-\eta_{\mu\nu}\epsilon^{\mu\nu}\left(\frac{1}{2}\partial^{\mu}\phi\partial_{\mu}\phi-V\left(\phi\right)\right)\right)\notag\\
 & =-i\kappa\dot{\phi}^{2}\epsilon^{00}=0,
 \label{eq:tmn}
\end{align}
the last diagram also do not contribute to the total differential decay width.
 Here, to evaluate \eqref{tmn}, we have used the gauge condition given in \eqref{conditiontraceless}, and a specific form of polarisation tensor based on the conditions given in \eqsref{conditiontransverse} - \refb{conditionorthogonal}, for which $\epsilon^{00}=0$.

To evaluate the differential decay width of the first two diagrams we will utilise the fact that the decaying scalar is a classical field.
Thus, for the first diagram we have
\begin{align}
M_{n}=-\phi\left(t\right)y_{\phi}\bar{v}\left(p_{3}\right)P_{R}\frac{\slashed{p_{1}}+\slashed{k}+M_{N_1}(t)}{\left(p_{1}+k\right)^{2}-M_{N_1}(t)^{2}}P_{R}\frac{i\kappa}{8}p_{2(\mu}\gamma_{\nu)}\bar{u}\left(p_{2}\right)\epsilon^{\mu\nu}\left(k\right)=\phi_{0}\mathcal{P}_{n}M_{\text{particle}}.
\label{eq:mn1}
\end{align}
Here $M_{\text{particle}}$ is the Feynman amplitude of the bremsstrahlung process assuming inflaton as a particle and is derived in \cite{Barman_2023}. 
Note that the Feynman amplitude for the second diagram is identical to  \eqref{mn1}.
The squared amplitude then evaluated as
\[
\sum_{\text{Pol}}\left|M\right|^{2}=\phi_{0}^{2}\left|\mathcal{P}_{n}\right|^{2}\sum_{\text{Pol}}\left|M_{\text{particle}}\right|^{2}=\frac{k\left(k-1\right)}{m_{\phi}^{2}}\rho_{\phi}\left|\mathcal{P}_{n}\right|^{2}\sum_{\text{Pol}}\left|M_{\text{particle}}\right|^{2}.
\]
To obtain the differential decay width needed in \eqref{BTEGWD}, we integrate over all the phase space variables of the fields and particles leaving out only the graviton energy 
\[
\frac{d\Gamma^{1\rightarrow3}}{dE_{k}}=\sum_{n}\int\frac{d\Pi_{\text{LIPS}}}{dE_{k}}\sum_{\text{Pol}}\left|M\right|^{2}.
\]
Here $n$ enumerates the oscillation modes of the inflaton. 
Averaging this over one oscillation period of the inflaton field gives us
\begin{equation}
\left\langle \frac{d\Gamma^{1\rightarrow3}}{dE_{k}}\right\rangle =\frac{\left(k+2\right)\left(k-1\right)}{64\pi^{3}}y_{\phi}^{2}\left(\frac{m_{\phi}\left(t\right)}{M_{P}}\right)^{2}\left(\frac{\omega\left(t\right)}{m_{\phi}\left(t\right)}\right)^{4}\sum_{n=1}^{\infty}n^{4}\left|\mathcal{P}_{n}\right|^{2}\frac{1-2x_{n}}{x_{n}}\left(2x_{n}\left(x_{n}-1\right)+1\right),\label{eq:decaywidthinflatongraviton}
\end{equation}
where $x_{n}=\frac{E_{k}}{nm_{\phi}}$.

\section{The $\phi\phi\rightarrow hh$ Annihilation Processes}
\label{sec:Ampl-GWS-spectrum}
Here, we will explore the effect of different Feynman diagrams in \figref{Feynman-diagrams-for-annihilation} on the total annihilation rate of inflaton producing pair of gravitons. 
Following the condition in \eqref{tmn}, the amplitudes of the first two diagrams are found to be zero as $M_{1,2}\propto T_{\mu\nu}\epsilon^{\mu\nu}$. 
The last two diagrams sum up to 
\begin{equation}
\sum_{s_{1},s_{2}}\left|M\right|^{2}=\sum_{n}\frac{\phi_{0}^{4}}{2M_{p}^{4}}\widetilde{m}_{\phi}^{4}\left|\mathcal{P}^{k}_{n}\right|^{2}=\sum_{n}\left|M_{n}\right|^{2}.
\end{equation}
Here we defined $\left|M_{n}\right|^{2}=\frac{\phi_{0}^{4}}{2M_{p}^{4}}\widetilde{m}_{\phi}^{4}\left|\mathcal{P}^{k}_{n}\right|^{2}$ as the contribution to the squared Feynman amplitude from each oscillation mode $n$ of the inflaton field and $\widetilde{m}_{\phi}^{2}=\frac{2m_{\phi}^{2}}{k\left(k-1\right)}$.
To get the rate of this annihilation reaction we perform phase space integral over all particles and fields
\begin{equation}
C\left[f_{\text{GW}}\right]=\left(\Delta N\right)\sum_{n}\int E_{\text{GW}}\left|M_{n}\right|^{2}D_{2}=\sum_{n}\int E_{\text{GW}}\frac{1}{8\pi^{2}}\frac{\left|k_{\text{GW}}\right|d\Omega}{E_{T}}\left|M_{n}\right|^{2}=\frac{\omega\left(t\right)\rho_{\phi}^{2}}{4\pi M_{p}^{4}}\Sigma^{k},\label{eq:collintann}
\end{equation}
where $E_{T}$ is the transverse energy, $\Sigma^{k}=\sum_{n}n\left|\mathcal{P}^{k}_{n}\right|^{2}$, $\mathcal{P}^{k}_{n}$ is $n$th Fourier coefficient of $\mathcal{P}^{k}$ and $\omega\left(t\right)$ is defined in \eqref{freqinflaton}.

\section{Spectrum of GW in the $\phi\phi\rightarrow hh$ Processes}
\label{sec:GWD-spectrum} 
The goal of this section is to evaluate the behaviour of the energy density of the gravitons  produced from inflaton annihilation as a function of graviton frequency.
To obtain such annihilation spectra, we need to solve the differential Boltzmann equation for annihilation process.
The Boltzmann transport equation is given by,
\[
L\left[f_{\rm GW}\right]=C\left[f_{\rm GW}\right],
\]
where the Liouvillian is 
\begin{equation}
L\left[f_{\rm GW}\right]=E\left(\dot{f}_{\rm GW}-Hp\frac{\partial f_{\rm GW}}{\partial p}\right)=HE\left(a\frac{\partial f_{\rm GW}}{\partial a}-E\frac{\partial f_{\rm GW}}{\partial E}\right),\label{eq:BTE}
\end{equation}
and the collision integral is 
\begin{align}
C\left[f_{\rm GW}\right] & =\left(\Delta N\right)\times\sum_{n}d\Pi_{h_{2}}^{LIPS}\left(2\pi\right)^{4}\delta^{4}\left(p_{n}-k_{1}-k_{2}\right)\left|M_{n}\right|^{2}=\frac{2\pi\rho_{\phi}^{2}}{M_{P}^{4}E_{1}}\left|\left(P^{k}\right)_{n}\right|^{2}\delta\left(E_{1}-\omega\left(a\right)\right)\label{eq:colintann}
\end{align}
Putting \eqsref{BTE} and \refb{colintann} together we get,
\begin{align}
\left(a\frac{\partial f_{\rm GW}}{\partial a}-E\frac{\partial f_{\rm GW}}{\partial E}\right) & =\frac{2\sqrt{3}\pi\rho_{\phi}\left(a_\text{end}\right)^{\frac{3}{2}}}{M_{P}^{3}E^{2}}\left|\left(P^{k}\right)_{n}\right|^{2}\left(\frac{a_\text{end}}{a}\right)^{\frac{9}{2}\left(1+w_{\phi}\right)}\delta\left(E-\omega\left(a\right)\right).\label{eq:diffBTEann}
\end{align}
Here we have assumed that the Hubble parameter $H$ is dominated by the inflaton and that the inflaton decay is negligible compared to the Hubble expansion at the beginning of the inflaton oscillation. 
This also allows us to obtain the frequency of the inflaton oscillation, which is same as the frequency of the emitted gravitons, as a function of the scale factor
\begin{equation}
\frac{\omega\left(a\right)}{\omega\left(a_\text{end}\right)}=\left(\frac{a}{a_\text{end}}\right)^{-3\left(1+w_{\phi}\right)\frac{k-2}{k}}.
\end{equation}
\eqref{diffBTEann} is easier to solve if we change variables to dimensionless quantities $y=\frac{a}{a_\text{end}}$, $x=\frac{E}{\omega\left(a_\text{end}\right)}$.
\begin{align*}
\left(y\frac{\partial}{\partial y}-x\frac{\partial}{\partial x}\right)f_{\rm GW} & =cx^{\beta}y^{\gamma}\delta\left(x-y^{\alpha}\right)=u\left(x,y\right),
\end{align*}
with appropriate powers $\alpha, \beta \text{ and } \gamma$. We can solve this PDE using method of characteristics. 
The characteristic equation is $\frac{dx}{x}=-\frac{dy}{y}$ which has the general solution $x =\frac{\xi}{y}$. 
This can be recognised as the redshift of energy with scale factor. 
Now we change variables $xy=\xi$ and $y=\eta$ to find the ODE on the characteristic lines as
\[
\eta\frac{\partial}{\partial\eta}f_{\rm GW}=u\left(\frac{\xi}{\eta},\eta\right).
\]
Integrating this ODE gives us the density function for the GW, 
\begin{align}
f_{\rm GW}\left(x,y\right) & =\int_{1}^{y}\frac{d\eta}{\eta}u\left(\frac{xy}{\eta},\eta\right)=c\left(xy\right)^{\beta}\int_{1}^{y}d\eta\eta^{\gamma-\beta-1}\delta\left(\frac{xy}{\eta}-\eta^{\alpha}\right)
\end{align}
Simplifying the $\delta$ function and evaluating the integral, we obtain,
\begin{equation}
f_{\rm GW}\left(x,y\right)=\frac{c}{2}\left|\frac{k+2}{k-4}\right|\left(xy\right)^{\frac{9}{k-4}}.\label{eq:solBTEGWS}
\end{equation}
We are interested in the energy spectrum and we note that it can be related to the density function from the definition
\begin{align*}
\rho_{\rm GW}\left(a\right)=\int\frac{d^{3}p}{\left(2\pi\right)^{3}}E\left(p\right)f_{\rm GW}\left(a,p\right) & =\int\frac{dE}{2\pi^{2}}E^{3}f_{\rm GW}\left(a,E\right).
\end{align*}
Consequently,
\begin{equation}
\frac{d\rho_{\rm GW}}{d\left(\ln E\right)}=E\frac{d\rho_{\rm GW}}{dE}=\frac{E^{4}}{2\pi^{2}}f_{\rm GW}\left(a,E\right)=\frac{\omega\left(a_\text{end}\right)^{4}x^{4}}{2\pi^{2}}f\left(x,y\right).\label{eq:energyGWS}
\end{equation}
Putting \eqsref{solBTEGWS}, \refb{energyGWS} together, we obtain the spectral behaviour of the energy density as
\begin{align*}
\frac{d\rho_{\rm GW}}{d\left(\ln E\right)} & =\frac{\sqrt{3}\omega\left(a_\text{end}\right)\rho_{\phi}\left(a_\text{end}\right)^{\frac{3}{2}}}{2\pi M_{P}^{3}}\left|\left(P^{k}\right)_{n}\right|^{2}\left|\frac{k+2}{k-4}\right|y^{\frac{9}{k-4}}x^{\frac{4k-7}{k-4}}\propto E^{\frac{4k-7}{k-4}}.
\end{align*}
Since the energy and frequency of gravitons are same in natural units
($\hbar=1$), $E=2\pi f$, we can recast this equation as
\begin{equation}
\frac{d\rho_{\rm GW}}{d\left(\ln f\right)}\propto f^{\frac{4k-7}{k-4}}.
\end{equation}
Requiring the total energy density to be consistent $\int\frac{d\rho_{\rm GW}}{d\left(\ln f\right)}d\left(\ln f\right)=\rho_{\rm GW}$
at the end of reheating $a=a_{ \rm RH}$ determines the proportionality
factor in terms of $\rho_{\rm GW}^{S}\left(a_{\rm RH}\right)$
\begin{equation}
\left.\frac{d\rho_{\rm GW}}{d\left(\ln f\right)}\right|_{a=a_{\rm RH}}=\left|\frac{4k-7}{k-4}\right|\frac{f^{\frac{4k-7}{k-4}}}{\left|f_{\rm RH}^{\frac{4k-7}{k-4}}-f_{e}^{\frac{4k-7}{k-4}}\right|}\rho_{\rm GW}^{S}\left(a_{\rm RH}\right).
\end{equation}
For $k=2$ this simplifies to 
\begin{equation}
\left.\frac{d\rho_{\rm GW}}{d\left(\ln f\right)}\right|_{a=a_{\rm RH}}=\frac{1}{2}\sqrt{\frac{f_{e}}{f}}\rho_{\rm GW}^{S}\left(a_{\rm RH}\right).\label{eq:spectrumRHGWS}
\end{equation}

\bibliographystyle{JCAP}
\bibliography{bibliography}

\end{document}